\documentclass[pra,superscriptaddress,twocolumn,longbibliography]{revtex4-1}
\usepackage{amsmath,mathrsfs}
\usepackage{graphicx}
\usepackage{rotating}
\usepackage{amssymb}

\usepackage[usenames, 
            dvipsnames
            ]{color}

\usepackage[colorlinks=true,
            linkcolor=Blue,
            filecolor=magenta,
            citecolor=Blue,
            urlcolor=Blue,
            pdftitle={Fermions in shaken square lattice},
            pdfauthor={A. Keles},
            bookmarks=true
            ]{hyperref}

\usepackage[sort&compress]{natbib}

\usepackage[numbered, 
            open
            ]{bookmark}

\usepackage[all]{hypcap}

\graphicspath{ {./figs/} }

\usepackage{multirow}

\usepackage{pifont}
\newcommand\uar{\ensuremath\uparrow}
\newcommand\dar{\ensuremath\downarrow}
\newcommand\dgr{\ensuremath^\dagger}
\newcommand\W{\ensuremath\mathcal{W}}
\newcommand\x{\ensuremath\mathbf{x}}
\newcommand\kb{\ensuremath\mathbf{k}}

\newcommand{\iu}{\mathrm{i}}
\newcommand{\ee}{\mathrm{e}}

\usepackage{bbold}
\newcommand\unity{\mathbb{1}}

\DeclareMathAlphabet{\mathpzc}{OT1}{pzc}{m}{it}
\newcommand\tb[1]{{#1}}

\newcommand\T{^\top}
\newcommand\n{\mathrm{n}}
\newcommand\s{\mathrm{S}}
\newcommand\p{\mathrm{P}}
\begin{document}
\title{Theory of interacting fermions in shaken square optical lattice}%{{{
\author{Ahmet Kele\c{s}}
\affiliation{Department of Physics and Astronomy,
  University of Pittsburgh, Pittsburgh, Pennsylvania 15260, USA}
\affiliation{Department of Physics and Astronomy,
  George Mason University, Fairfax, Virginia 22030, USA}
\author{Erhai Zhao}
\affiliation{Department of Physics and Astronomy,
  George Mason University, Fairfax, Virginia 22030, USA}
\author{W. Vincent Liu}
\affiliation{Department of Physics and Astronomy, 
  University of Pittsburgh, Pittsburgh, Pennsylvania 15260, USA}

\begin{abstract}  
  We develop a theory of weakly interacting fermionic atoms in shaken optical
  lattices based on the orbital mixing in the presence of time-periodic
  modulations.  Specifically, we focus on fermionic atoms in circularly shaken
  square lattice with near resonance frequencies, i.e., tuned close to the
  energy separation between $s$-band and the $p$-bands.  First, we derive a
  time-independent four-band effective Hamiltonian in the non-interacting
  limit.  Diagonalization of the effective Hamiltonian yields a quasi-energy
  spectrum consistent with the full numerical Floquet solution that includes
  all higher bands.  In particular, we find that the hybridized $s$-band
  develops multiple minima and therefore non-trivial Fermi surfaces at
  different fillings. We then obtain the effective interactions for atoms in
  the hybridized $s$-band analytically and show that they acquire momentum
  dependence on the Fermi surface even though the bare interaction is
  contact-like.  We apply the theory to find the phase diagram of fermions
  with weak attractive interactions and demonstrate that the pairing symmetry
  is $s+d$-wave.  Our theory is valid for a range of shaking frequencies near
  resonance, and it can be generalized to other phases of interacting fermions
  in shaken lattices.
\end{abstract} 

\maketitle
%}}}
\section{Introduction}%{{{
\label{sec:intro}

Manipulating quantum many-body systems by time-periodic driving, known as
Floquet engineering, has recently emerged as a powerful way to experimentally
control the band structure of ultracold atoms in optical lattices. One
particular useful approach is lattice shaking, namely moving the entire
lattice along a certain prescribed periodic trajectory in space by tuning the
phases of interfering laser beams that give rise to the optical lattice
potential. With a properly designed shaking protocol, the time-dependent
many-body system may resemble a time-independent system described by an
effective Hamiltonian with desired properties. 

Floquet engineering by lattice shaking has been successfully demonstrated in a
number of experiments. Dynamical control of tunneling and band flattening in
optical lattice was reported in
\cite{PhysRevLett.81.5093,PhysRevLett.99.220403}.  Superfluid-Mott insulator
transition induced by shaking has also been observed \cite{Zenesini2009}.
Lattice shaking can give rise to artificial gauge fields for neutral atoms,
complementary to other approaches based on for example the Raman coupling
scheme or laser insisted tunneling (for review, see for instance
\cite{Holthaus2015a,Goldman2014a,eckardt2016atomic}).  
Along this line, increasingly sophisticated shaking schemes were implemented
to realize, for instance, frustrated magnetism on triangular
lattice\cite{Struck2011}, tunable gauge fields including staggered flux
\cite{Struck2012,Hauke2012b}, the $XY$-spin model \cite{Struck2013}, 
and the Haldane model on honeycomb lattice \cite{Jotzu2014}.  Shaken
one-dimensional lattice with mixed Bloch bands and the resulting spectrum with
double minima was used to simulate ferromagnetism \cite{Parker2013} and to
test the universality relations near quantum phase transitions
\cite{Clark2016}.

Stimulated by these experiments, theoretical work has explored and clarified
various regimes of shaking. Shaken lattice is intrinsically a multi-band
problem, where the relevant energy scales are the band widths, energy gaps,
and the shaking frequency $\omega$. For deep optical lattices and when
$\omega$ is much larger than the width of the lowest $s$-band (but much
smaller than the excitation gap), a single band picture is possible and the
system can be described by a static Hamiltonian with an effective tunneling
amplitude tunable in both magnitude and sign \cite{PhysRevLett.95.260404}. In
this frequency regime, if the shaking protocol is adjusted to have certain
symmetries, artificial gauge fields can be realized
\cite{PhysRevLett.94.086803,Hauke2012b,PhysRevX.4.031027}.  An extensively
studied shaken system is the two-dimensional hexagonal lattice in the tight
binding limit. When the shaking frequency matches the energy difference
between the sublattice sites, unconventional pairing is predicted to occur for
fermions \cite{Zhang2015a}.  Tuning the shaking frequency to match the
band-gap at the Brillouin zone center leads to the so-called moat band
\cite{Sedrakyan2015}. For larger $\omega$, the higher Bloch bands have to be
incorporated and band mixing may drastically modify the band structure. For
example, when $\omega$ is tuned to half of the energy gap between $s$-band and
$p$-band, two-photon processes dominate and give rise to topologically
nontrivial bands in both one dimension \cite{Zheng2014} and two dimensions
\cite{Zhang2014}. Finally, when $\omega$ is on the order of the gap between
the $s$- and $p$-bands, single photon processes dominate and lead to
dispersion spectra with multiple minima implying interesting many-body
phenomena \cite{Zheng2015}.

In contrast to the single particle properties, understanding the interaction
effects for fermions in shaken lattices remains a theoretical challenge. The
problem is complicated by the time dependence and the hybridization of
multiple Bloch bands. A key point is that the effective interactions between
the atoms are modified by shaking and must be derived and analyzed along with
the band structures. 
In this paper, we formulate an effective theory to capture the essential
physics of interacting fermions on shaken lattices.  For concreteness, we will
focus on the two-dimensional square lattice with circular shaking and near
resonance frequencies, i.e., with $\omega$ tuned close to the energy
separation between $s$- and $p$-bands at the Brillouin zone center. Our theory
is not restricted to the tight binding limit and various values of the optical
lattice depth will be considered.

This paper is organized as follows. First we describe the problem and
summarize our main results in Section \ref{sec:formalism}. The single particle
spectrum for the shaken lattice is solved numerically in
Sec.~\ref{sec:numerics} using the standard Floquet analysis. The main goal
here is to obtain the quasienergy spectrum accurately, which will be used to
benchmark our approximation schemes. For example, the hybridized $s$-band is
shown to develop four minima as opposed to a single minimum at the Brillouin
zone center. 
By comparing the numerical spectrum with simple folding construction, we show
that it is sufficient to keep only a few lowest Bloch bands. This observation
motivates our subsequent analytical theory.
In Sec.~\ref{sec:effective_model}, we derive a four-band effective Hamiltonian
$H_\mathrm{eff}$ for the single-particle shaken system by using the rotating
wave approximation (RWA) that is consistent with the numerical Floquet
solution.  Diagonalizing $H_\mathrm{eff}$, we determine the non-trivial Fermi
surface geometries of fermionic atoms populating the hybridized $s$-band as
function of lattice filling fraction.
Next, we take interactions into account in Sec.\ref{sec:interactions} and
derive the effective interaction $V_\mathrm{eff}(\mathbf{k},\mathbf{k}')$ for
fermions on the Fermi surface of the hybridized $s$-band. 
%This reduces the complicated, multi-orbital interaction terms to a
%simple form which can be computed analytically. 
%Our effective theory
%has a remarkable result; the complicated, multi-orbital interaction term 
%reduces to an unexpected simple form, even after all analytical manipulations
%such as RWA. 
In particular, we show that it develops interesting momentum
dependence and no longer point-like.
In Sec.~\ref{sec:pairing}, we apply the effective model to fermions with weak
attractive interactions in circularly shaken square lattice. We investigate
the pairing symmetry and transition temperature for different filling
fractions. We show that the order parameter can have $s+d$-wave symmetry.
The theoretical framework is applied in Sec.~\ref{sec:resonant} to study
red-detuned near-resonance shaking, where the bands directly overlap.  We
conclude with remarks on the implications of our work in
Sec.~\ref{sec:conclude}.

%}}}
\section{The problem and main results}%{{{
\label{sec:formalism}

We are interested in formulating a theory of interacting fermionic ultracold atoms in
an optical lattice potential that varies periodically in time. Consider for
example a square optical lattice given by the potential
\begin{equation}
    V_\mathrm{lat}(\x) = V_0\left[ \cos(\frac{2\pi x}{\lambda_L}) +
    \cos(\frac{2\pi y}{\lambda_L})\right].
  \label{eq:lattice_potential}
\end{equation}
Here $\x=(x,y)$ is the space coordinate in two dimensions (2D), $\lambda_L$ is
the wavelength of the lasers forming the optical lattice, $V_0$ is the lattice
depth and we assume tight confinement in the $z$-direction.  For a shaken
optical lattice, the lattice potential $V_\mathrm{lat}(\x)$ is replaced by
$V_\mathrm{lat}(\x+\x_0(\tau))$ and becomes a periodic function of time
$\tau$, since the origin of the lattice, $\x_0(\tau)$, moves along a
prescribed loop in space and returns to its starting point after one shaking
period $T$, 
\begin{equation}
\x_0(\tau)=\x_0(\tau+T).
\end{equation}
Different choices of $\x_0(\tau)$ are referred to as different shaking
protocols. For example, a general shaking protocol corresponds to the
choice
\begin{equation}
    \mathbf{x}_0(\tau) = s_0[\sin\omega\tau, \sin(\omega\tau+\vartheta)],
\end{equation}
where $s_0$ is the amplitude of shaking, $\omega\equiv2\pi/T$ is the shaking
frequency and $\vartheta$ is the phase difference between $x$ and $y$
directions. In this paper we take $\vartheta=\pi/2$ which correspondd to
circular shaking.
 
The many-body system of fermionic atoms loaded in such lattices is described
by an action $\mathcal{S} = \int d\tau [\mathscr{L}_0+\mathscr{L}_1]$ where
the single particle part of the Lagrangian $\mathscr{L}_0$ has the form
\begin{align}
  \mathscr{L}_0 & = \int d\x 
  \psi_\sigma\dgr(\x,\tau) 
  \left[ 
    \iu\partial_\tau - H_0(\x,\tau)
  \right]
  \psi_\sigma(\x,\tau).
\end{align}
Here $\sigma=\uar,\dar$ is the spin index and we take $\hbar=1$. For shaken
optical lattices,
\begin{equation}
  H_0(\x,\tau) = \frac{\mathbf{p}^2}{2m} 
  + V_\mathrm{lat}(\x+\x_0(\tau)) 
  \label{eq:H_shaken}
\end{equation}
with $\mathbf{p}=-i\nabla$. Following the discussion above regarding
$V_\mathrm{lat}$, the single particle Hamiltonian $H_0$ is periodic both in
time $H_0(\x,\tau)=H_0(\x,\tau+T)$ and in space
$H_0(\x,\tau)=H_0(\x+\mathbf{R}_i,\tau)$ where $\mathbf{R}_i$ are the 
lattice vectors and $T$ is the shaking period.  Note that the single-particle
time-dependent Schrodinger equation can be obtained as $\iu\partial_\tau\psi =
H_0\psi$.

The fermionic field operators obey the equal-time anti-commutation relation
\begin{equation}
  \left\{ 
    \psi_\sigma(\x,\tau), 
    \psi_{\sigma'}\dgr(\x',\tau)
  \right\} 
  = \delta_{\sigma\sigma'}\delta(\x-\x').
\end{equation}
For a dilute gas of ultracold alkali atoms, the interactions are local both
in time and space such that the two particle interaction potential takes the
form
\begin{equation}
  U(\x-\x',\tau-\tau')=g
  \delta(\x-\x')
  \delta(\tau-\tau').
\end{equation}
Here the interaction strength $g$ is related to the low energy s-wave
scattering length. The interaction part of the Lagrangian,
$\mathscr{L}_1$, can be written as
\begin{align}
  \mathscr{L}_1 & = g\int d\x 
  \psi_\uar\dgr(\x,\tau)
  \psi_\dar\dgr(\x,\tau)
  \psi_\dar(\x,\tau)
  \psi_\uar(\x,\tau).
  \label{eq:interaction}
\end{align}

Our strategy is to study the single particle physics of $\mathscr{L}_0$ first
and then incorporate interaction $\mathscr{L}_1$ later. Even without
$\mathscr{L}_1$, it is a challenge to analyze the time-dependent Hamiltonian
$H_0(\x,\tau)$ directly.  To make progress, we recast $H_0$ in slightly
different forms that are convenient for subsequent analytic or numeric
treatments.  This is achieved by performing a gauge transformation to co-moving
frame, $\psi'=e^{-i\x_0\cdot\mathbf{p}}\psi$, in which $H_0$ becomes
\begin{equation}
  H'_0 = \frac{\mathbf{p}^2}{2m} + V_\text{lat}(\x)
  + \dot{\mathbf{x}}_0(\tau) \cdot\mathbf{p}
  \label{eq:H_driven}
\end{equation}
with $\dot{\mathbf{x}}_0(\tau)=\partial\mathbf{x}_0/\partial\tau$. Note the
result is valid for arbitrary shaking protocol. In this form, the driving
appears as a time-dependent perturbation $\dot{\mathbf{x}}_0(\tau)
\cdot\mathbf{p}$ to the static problem $\mathbf{p}^2/{2m} + V_\text{lat}(\x)$.
The last term in Eq. \eqref{eq:H_driven} can be combined with the first term
by completing the square. The resulting time-dependent term proportional to
$\dot{\mathbf{x}}^2_0$ can be removed via another gauge transformation
\cite{Holthaus2015a}. In this case, Hamiltonian becomes
\begin{equation}
    H''_0 = \frac{[\mathbf{p}+m\dot{\mathbf{x}}_0(\tau)]^2}{2m} + 
    V_\mathrm{lat}(\mathbf{x}). 
    \label{eq:H_gaugefield}
\end{equation}
This form suggests that the lattice shaking is equivalent to the presence of a
time-dependent vector potential 
\begin{equation}
\mathbf{A}=-m\dot{\mathbf{x}}_0
\end{equation}
and the corresponding force field is given by $\mathbf{E}(\tau)=-\partial
\mathbf{A}(\tau)/\partial\tau=m\ddot{\mathbf{x}}_0$.  The presence of a vector
field $\mathbf{A}$ may drastically modify the band dispersions.
We stress that the Hamiltonians given in Eqs.~\eqref{eq:H_shaken},
\eqref{eq:H_driven} and \eqref{eq:H_gaugefield} are equivalent. For the
numerics in section \ref{sec:numerics}, we use Eq.~\eqref{eq:H_gaugefield}
which is consistent with Ref.~\cite{Holthaus2015a}.  For analytical
manipulations in section \ref{sec:effective_model} and after, we use
Eq.~\eqref{eq:H_driven} instead to treat time dependence separately.

% In this work, we focus on the example of spin 1/2 fermions with contact
% interactions in circularly shaken square optical lattice.
% And we are mostly interested in the regime of near resonance shaking where
% the shaking frequency $\omega$ is comparable to the energy separation of the 
% lowest $s$-band and the doubly degenerate $p$-bands. Thus band mixing 
% (hybridization) is expected to be strong and a multi-band description is necessary. 
% It is however unclear a priori whether a simple effective Hamiltonian can capture the
% essential physics of lattice shaking, and what are the consequences of weak interactions
% on the shaken lattice. For example, it turns out that it is necessary to take into
% account the lowest lying $d$-band.

Our key results for the readers who wish to skip the technical details are
summarized as follows.
For near resonance shaking, the quasienergy spectrum can be captured by a
static, four-band effective Hamiltonian $H_\mathrm{eff}$ given in
Eq.~\eqref{eq:rwa_hamiltonian} where the band mixing is described by the
off-diagonal matrix elements.
$H_\mathrm{eff}$ can be diagonalized analytically to yield the dispersion of
the hybridized bands.  For example, the hybridized $s$-band dispersion is
given by $\epsilon_1(\kb)=\epsilon_{++}(\kb)$ in
Eq.~\eqref{eq:effective_dispersion}. The effective interactions for fermions
on the hybridized $s$-band is given by $V_\mathrm{eff}(\kb,\kb')$ in
Eq.~\eqref{eq:full_interaction}. 
Based on the effective band structure and interactions, we solve the pairing
problem of fermions with weak attractive interaction in the hybridized $s$-band to get
the phase diagram and the symmetry of the order parameter as
shown in Fig.~\ref{fig:phase_diagram}.

%}}}
\section{Quasienergy spectrum from Floquet analysis}%{{{
\label{sec:numerics}

We first numerically calculate the single particle spectrum of time periodic
Hamiltonian $H''_0$ in Eq.~\eqref{eq:H_gaugefield}. This problem is previously
considered in Refs.~\cite{Miao2015} and \cite{Po2015} for bosonic systems
using Wannier expansion up to three and four orbitals, respectively.  Here we
adopt an approach based on Bloch expansion that can include all higher bands
to desired numerical accuracy. 

Floquet operator $\mathscr{U}(T)$  is defined as the time evolution operator
over one shaking period $T$. It can be written as the following time ordered
exponential,
\begin{align}
    \mathscr{U}(T) 
    &= \mathcal{T}\exp\left\{-\iu \int_0^T d\tau H''_0(\tau) 
    \right\}.
\end{align}
We expand the wave functions using the Bloch theorem 
\begin{equation}
\psi_\mathbf{k}(\mathbf{x},\tau)=e^{i\mathbf{k}\cdot
  \mathbf{x}}\sum_{\mathbf{G}}\Psi_\mathbf{k}(\mathbf{G},\tau)
\ee^{\iu\mathbf{G}\cdot\mathbf{x}}
\end{equation}
where $\Psi_\mathbf{k}(\mathbf{G},\tau)$ are expansion coefficients for
crystal momentum $\kb\equiv(k_x,k_y)$ and reciprocal lattice vectors are given
by $\mathbf{G}=\frac{2\pi}{\lambda_L}\boldsymbol{\ell}$ with intergers
$\boldsymbol{\ell}=(\ell_x,\ell_y)$. In numerical calculations, we take
$\ell_{x,y}=-N_b,\cdots,N_b$ where the cutoff $N_b=4$ corresponds to the
inclusion of $(2N_b+1)^2=81$ orbitals. We checked that increasing $N_b$
further does not change the results. To express the resulting Hamiltonian
matrix in a simple form, we measure lattice momentum in units of the recoil
momentum $k_L=\pi/\lambda_L$, energy and shaking frequency in units of the
recoil energy $E_R=k_L^2/2m$: $k\rightarrow k/k_L$, $H_0\rightarrow H_0/E_R$
and $\omega/E_R\rightarrow\omega$. Then Eq.~\eqref{eq:H_gaugefield} becomes
\begin{equation}
  H''_0(\boldsymbol{\ell},\boldsymbol{\ell}';\tau) = 
  \left[ 
    \kb + 2\boldsymbol{\ell} + \mathbf{A}(\tau) 
  \right]^2\delta_{\boldsymbol{\ell},\boldsymbol{\ell}'}+
  V_{\boldsymbol{\ell},\boldsymbol{\ell}'},
  \label{eq:H_bloch_expansion}
\end{equation}
where the $\kb$ dependence of $H''_0$  is suppressed for brevity. For the
square optical lattice potential given in Eq.~\eqref{eq:lattice_potential},
the matrix elements can be calculated as
$V_{\boldsymbol{\ell},\boldsymbol{\ell}'} =
(V_0/2E_R)\left[\delta_{\ell_x',\ell_x+1}+\delta_{\ell_x',\ell_x-1}\right]
\delta_{\ell_y',\ell_y}+(x\leftrightarrow y)$ and the vector potential coming
  from shaking has the form 
\begin{equation}
 \mathbf{A}(\tau) = \beta \left[ \cos\tau,
   \cos(\tau+\vartheta)\right]    
\end{equation}
where the dimensionless shaking strength is defined as 
\begin{equation}
\beta = (\omega/E_R)(s_0k_L). 
\end{equation}
The time ordered product is calculated by dividing the time evolution into
many small slices $\{\tau_i\}$,
\begin{align}
    \mathscr{U}(T) 
    =\prod_{\tau_i=0}^{2\pi}
    \exp\left\{-\frac{\iu}{\omega} 
      H''_0(\boldsymbol{\ell},\boldsymbol{\ell}';\tau_i) \right\}.
\end{align}
For a given $\kb$ and discrete $\tau_i$, we compute the corresponding matrix
exponentials in Eq.~\eqref{eq:H_bloch_expansion}. Then by taking their product,
we obtain the Floquet operator.  The eigenvalues of the Floquet operator,
$\mathscr{U}(T) v_n(\kb) =\Lambda_n(\kb) v_n(\kb)$  has
the form $\Lambda_n(\kb)=\ee^{-\iu T\mathscr{E}_n(\kb)}$ which defines the quasi-energy
spectrum by the relation
\begin{equation}
    \mathscr{E}_n(\kb) = -\frac{1}{T}\mathrm{Im}\log
    \Lambda_n(\kb) 
    \label{eq:floquet_energy}
\end{equation}
where $n$ is the Floquet band index.  Note that the quasienergy spectrum is
periodic, i.e., an energy level at $\mathscr{E}_n(\kb)$ is identical to
$\mathscr{E}_n(\kb)+\omega$. The region $\mathscr{E}_n\in [0, \omega]$ is
called the quasienergy Brillouin zone (QeBZ), analogous to the quasimomentum
Brillouin zone of a solid. 

\begin{figure}[t]
  \centering
  \includegraphics[width=0.5\textwidth]{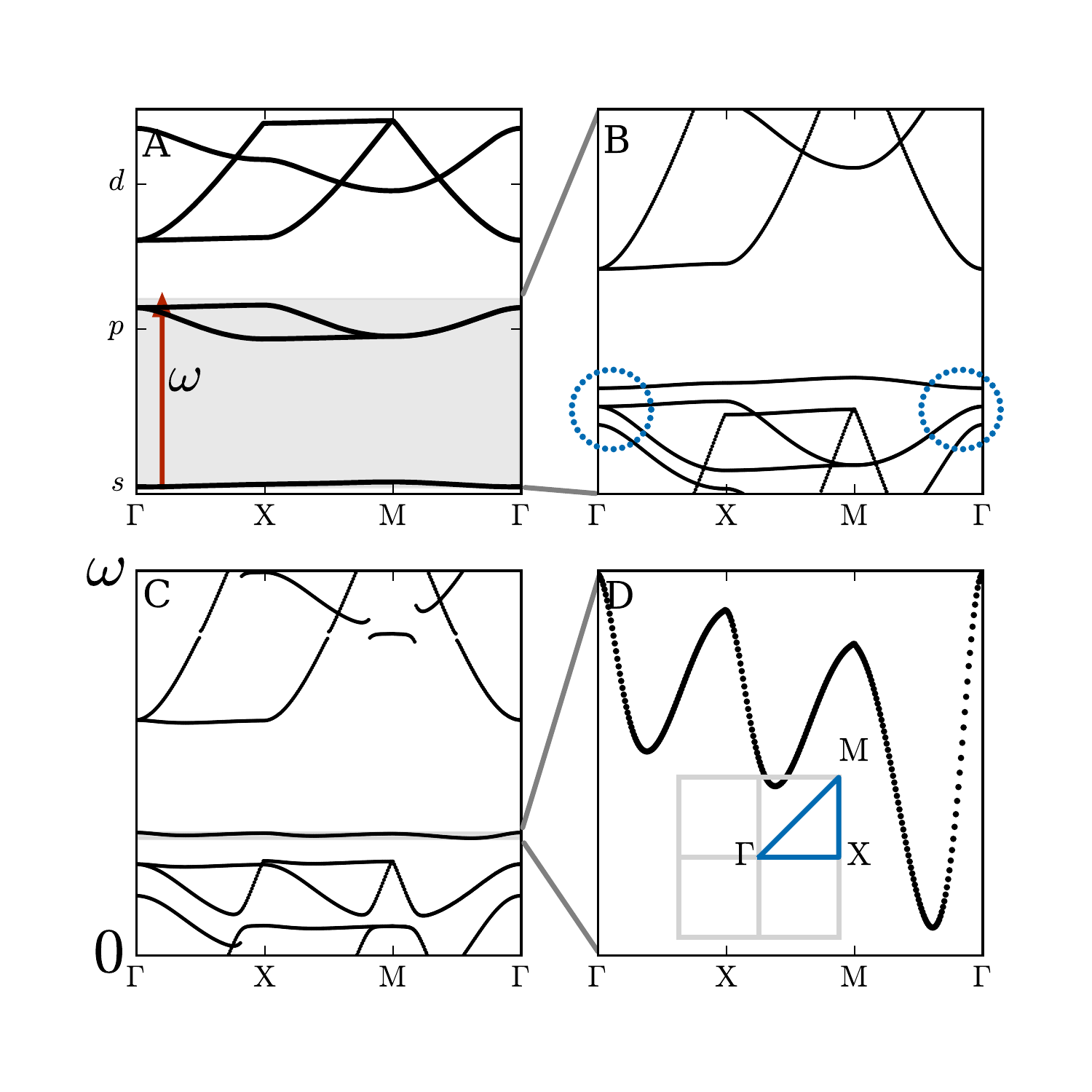}
  \includegraphics[width=0.5\textwidth]{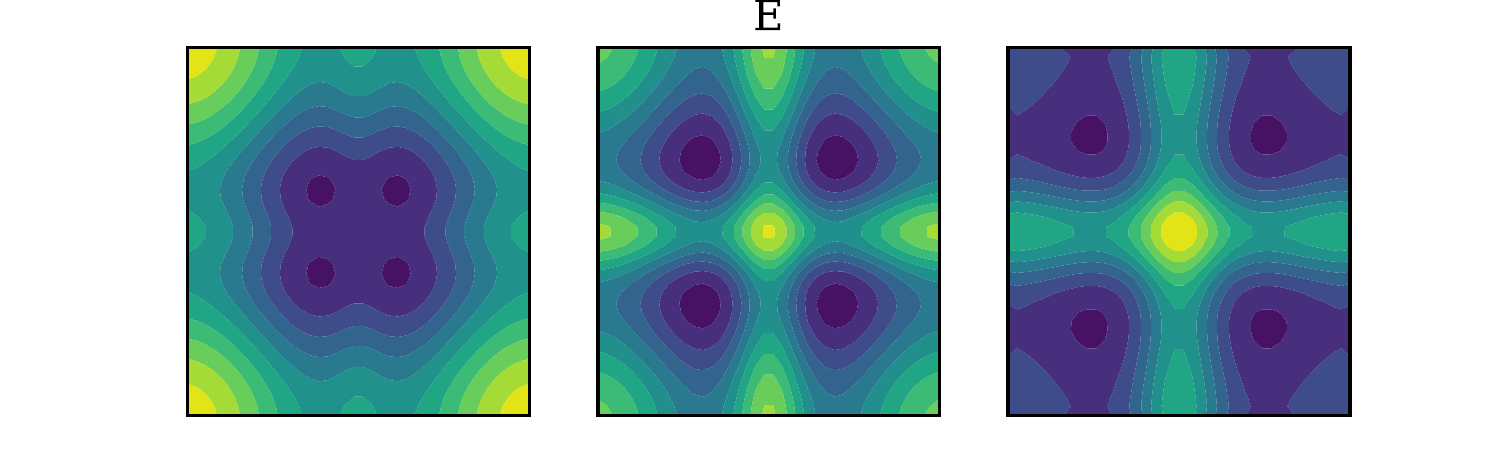}
  \caption{(Color online) Band structure of shaken square lattice from the
    numerical Floquet analysis.  \textbf{A.} The lowest few bands for lattice
    depth $V_0/E_R=5$ without shaking. The shaking frequency $\omega$ is shown
    with (red) arrow. It is fixed at $\omega=1.05\Omega$ where $\Omega=\max
    E_p-\min E_s$. The first quasienergy Brillouin zone (QeBZ) is indicated by
    the gray shaded region.  \textbf{B.} The bands folded into QeBZ.  The
    (blue) circles indicate regions where strong hybridization is expected to
    occur between $s$, $p_x$, $p_y$ and $d_{xy}$-bands. \textbf{C.} The
    quasienergy spectrum of circularly shaken square lattice with shaking
    amplitude $\beta=0.1$. For clarity only six relevant eigenvalues of the
    Floquet solution are shown. Band mixing leads to fine features not
    captured by the folding construction.  \textbf{D.} The zoomed details of
    the hybridized $s$-band. The inset shows the high symmetry points inside
    the quasi-momentum Brillouin zone.  \textbf{E.} Contour plot of the
    dispersion $\mathscr{E}_s(\kb)$ of the hybridized $s$-band in the
    Brillouin zone obtained from the numerical Floquet solution for shaking
    amplitudes $\beta=0.1$ (left), $\beta=0.2$ (middle) and $\beta=0.3$
    (right). Dark (Blue) regions correspond to low energies, while bright
    (yellow) regions correspond to high energies.  Horizontal and vertical
    axes are $k_x$ and  $k_y$ in units of recoil momentum $k_L$.  }
  \label{fig:shaken_band}
\end{figure}

The quasienergy band structure depends on two independent energy scales: the
lattice depth in terms of recoil energy $V_0/E_R$  and the shaking frequency
$\omega/E_R$ which couples different energy sectors with a strength on the
order of $\beta$. We have explored several regimes in our numerics and found
that one of the most interesting regimes is near resonance shaking, i.e., when
$\omega$ is comparable to $\Omega$ defined by
\begin{equation}
    \Omega\equiv \max E_p-\min E_s,
\end{equation}
where the bottom of the $s$-band is $\min E_s$ and the top of
two-fold degenerate the $p$-bands is $\max E_p$. For such frequencies, shaking
strongly couples the $s$- and $p$-bands.  This is illustrated in
Fig.~\ref{fig:shaken_band}A, where $\omega$ is set to $\omega=1.05\Omega$. The
bands in Fig.~\ref{fig:shaken_band}A are labeled using standard notation,
e.g.,  the $s$, $p_x$, $p_y$ and $d$-bands, in line with Refs.
\cite{Po2015,Miao2015}. {Note that the
orbital symmetries like $s$, $p_{x,y}$ and $d$ are mixed in the shaken system
so formally it is not possible to label bands by their symmetries.
However, mixed orbitals still retain a dominant symmetry after shaking as
long as shaking is not perturbatively strong. Based on this reasoning, we call the
bands after shaking hybridized $s$-band, hybridized $p$-bands etc.}

The gross features of the quasienergy spectrum of the shaken lattice can be
captured by a simple folding construction: take the static bands given in
Fig.~\ref{fig:shaken_band}A and fold them into the same QeBZ,
$\mathscr{E}_n(\kb)\rightarrow \mathrm{mod}_\omega[E_n(\kb)]$, one then
obtains a rough caricature of the quasienergy spectrum shown in
Fig.~\ref{fig:shaken_band}B. We observe that, after folding, the $s$, $p_x$,
$p_y$ and one of the $d$-bands, $d_{xy}$, come close in energy in regions
indicated by the blue circles around the $\Gamma$ point. For the shaking
frequency chosen here, even though there is no direct level crossing inside
these circles and the $s$-band seems to be isolated, the mutual influence of
these bands turns out to be important.  There are however level crossings
between higher bands, e.g., between the $p$- and $d$-bands. 

Fig.~\ref{fig:shaken_band}C shows the full numerical solution of the
quasienergy band structure.  By comparing it to Fig.~\ref{fig:shaken_band}B,
we can identify the relatively flat band within the small gray region as the
hybridized $s$-band, which is plotted separately for clarity in
Fig.~\ref{fig:shaken_band}D  with a finer energy resolution. We notice that
the level repulsion between the $s$-band and $p$-bands, particularly around
the $\Gamma$ point, pushes up the bottom of the $s$-band located at
the $\Gamma$ point. As a result, the minimum of the $s$-band moves away from
$\Gamma$ to four $\mathbf{k}$ points on the diagonal $\Gamma-\mathrm{M}$ line.  The four
minima can also be seen in the full dispersion shown in
Fig.~\ref{fig:shaken_band}E for three different shaking amplitudes.  It
clearly demonstrates that the $s$-band is strongly modified by lattice
shaking. Similar effect has recently  been observed in experiments with bosons
in one-dimensional shaken lattice in Ref.~\cite{Parker2013}.  

There are also other dramatic consequences of shaking to the quasienergy spectrum.
For example, many of the level crossings of the $p$-bands 
and $d$-bands in Fig.~\ref{fig:shaken_band}B become avoided
crossings in Fig.~\ref{fig:shaken_band}C. Moreover, the $d$-bands
are also modified by coupling to higher bands 
such as the $f$-bands (not shown). Hereafter we are mainly interested in 
what happens to the $s$-band in the presence of non-perturbative shaking. 
For this purpose, our numerical findings here suggest that it is sufficient
to consider the lowest four orbitals; $s$, $p_x$, $p_y$, and $d_{xy}$. 

%}}}
\section{Effective Four-band model}%{{{
\label{sec:effective_model}

In this section, we derive analytically an effective four-band Hamiltonian
that can accurately describe the quasienergy spectrum up to the $d_{xy}$ band,
consistent with the numerical Floquet solution.
Based on previous section, a truncation up to four lowest Bloch bands is sufficient to
understand the interplay of multiple orbitals in the shaking problem, as far
as the modifications to the $s$-band and $p$-bands are concerned. The
effective Hamiltonian will serve  as the basis to study interaction effects in
the next section. 

%To keep the time dependent driving as a separate perturbation, 

To analyze the Hamiltonian $H'_0$ in Eq.~\eqref{eq:H_driven}, we first carry
out Wannier expansion for the wavefunction,
\begin{equation}
    \psi(\x,\tau) = \sum_{i,n} \psi_{i,n}(\tau)
    \mathcal{W}_n(\x-\mathbf{R}_i).
    \label{eq:wannier_expansion}
\end{equation}
Here $\mathcal{W}_n(\x-\mathbf{R}_i)$ are Wannier functions localized at site
$i$ for the static square optical lattice.  We will truncate the orbital index
$n$ and only keep four orbitals, $n=s, p_x, p_y, d_{xy}$.  The truncation
error can be assessed by comparing to the full numerical results of the
previous section.
In this basis, the Schr\"odinger equation becomes
\begin{equation}
  \iu \partial_\tau \psi_{i,n}(\tau)  = 
    \sum_{i',n'} [H'_0(\tau)]_{ii',nn'}
    \psi_{i',n'}(\tau),
    \label{eq:H_wannier_expansion}
\end{equation}
where $H'_0$ is a $4\times 4$ matrix in orbital space.  Since $H'_0$ naturally
splits into a static part $H_\mathrm{static}=\mathbf{p}^2/{2m} + V_\text{lat}(\x)$
describing the unshaken lattice and a time-dependent part
$V(\tau)=\dot{\mathbf{x}}_0(\tau) \cdot\mathbf{p}$ describing shaking, we
evaluate their matrix elements in turn as follows.

The familiar $H_\mathrm{static}$ contains onsite and the nearest-neighbor hopping terms
\begin{align}
[H_\mathrm{static}]_{ii'}&=\delta_{i,i'} \mathrm{Diag}(\tb{e}_s, \tb{e}_p,
\tb{e}_p,
\tb{e}_d) \nonumber \\
&+\sum_\pm \delta_{i,i'\pm\hat x}
 \mathrm{Diag}(\tb{t}_s, \tb{t}_{p}, \tb{t}_{s}, \tb{t}_p)\nonumber\\
&+\sum_\pm\delta_{i,i'\pm\hat y}\mathrm{Diag}(\tb{t}_s, \tb{t}_{s}, \tb{t}_{p}, \tb{t}_p).
\end{align}
The onsite energy for each band is defined as 
\begin{equation}
\tb{e}_n=\int d\mathbf{x}
\mathcal{W}^*_n(\mathbf{x})H^s_0(\x)\mathcal{W}_n(\mathbf{x}),
\end{equation}
and the hopping integrals for the $s$- and $p$-orbitals are given by
\begin{align}
\tb{t}_s &=
\int d\mathbf{x} \mathcal{W}^*_s(\mathbf{x}) H^s_0(\x) \mathcal{W}_s(\mathbf{x}+ \hat x), 
\nonumber\\
\tb{t}_p &= \int d\mathbf{x} \mathcal{W}^*_{p_x}(\mathbf{x})
H^s_0(\x)
\mathcal{W}_{p_x}(\mathbf{x}+ \hat x),
\end{align}
where we have taken the lattice spacing to be one. 

Similarly, we can split the time periodic shaking term $V(\tau)$ into onsite
($V_0$) and nearest neighbor ($V_1$) coupling terms,
\begin{align}
[V(\tau)]_{ii'}&=\delta_{i,i'}
\left[ 
  a_x(\tau)V_0^x + a_y(\tau)V_0^y 
\right] \nonumber \\
&+ \sum_\pm\delta_{i,i'\pm\hat x}a_x(\tau)V_1^x \nonumber \\
&+ \sum_\pm\delta_{i,i'\pm\hat y}a_y(\tau)V_1^y. \label{vtau}
\end{align}
Here $\mathbf{a}(\tau)=\partial_\tau \mathbf{x}_0(\tau)$, i.e.,
\begin{equation}
a_x(\tau)=s_0\omega \cos (\omega\tau),\;\;  a_y(\tau)=s_0\omega \cos (\omega\tau+\vartheta).
\end{equation}
The symmetry of $V(\tau)$ dictates that for the onsite terms, only $s-p_{x,y}$
and $p_{x,y}-d_{xy}$ couplings are allowed,
\begin{align}
  % onsite ax shaking
  V_0^x &= 
  \left[
    \begin{array}{cccc}
      0          & \tb{d}_0 &            & \\
      \tb{d}_0^* & 0        &            & \\
      &          & 0          & \tb{d}_0  \\
      &          & \tb{d}_0^* & 0
    \end{array}
  \right],\;\;\;
  % onsite ay shaking
  V_0^y = 
  \left[
    \begin{array}{cccc}
      &            & \tb{d}_0 & 0  \\
      &            & 0        & \tb{d}_0  \\
      \tb{d}_0^* & 0          &          & \\
      0          & \tb{d}_0^* &          &
    \end{array}
  \right].
\end{align}
Such shaking induced band mixing is characterized by the coupling strength
\begin{equation}
\tb{d}_0=\int d\mathbf{x}\mathcal{W}^*_{s}(\mathbf{x})\hat
p_x\mathcal{W}_{p_x}(\mathbf{x}).
\end{equation}
The nearest neighbor coupling terms have a similar
matrix structure, 
\begin{align}
  % x hopping by ax shaking
  V_1^x &= 
  \left[
    \begin{array}{cccc}
     \iu \tb{t}_s'     & \tb{d}_1 &               & \\
     \tb{d}_1^* & \iu \tb{t}_p'           &          & \\
                 &        & \iu \tb{t}_s'         & \tb{d}_1  \\
                 &        & \tb{d}_1^* & \iu \tb{t}_p'
    \end{array}
    \right],\;\;\;
  V_1^y = 
  \left[
    \begin{array}{cccc}
      \iu \tb{t}_s'&      &   \tb{d}_1   &  0  \\
     &    \iu \tb{t}_s'     &   &  \tb{d}_1  \\
     \tb{d}_1^*   &   0   & \iu \tb{t}_p'     &    \\
     0   &   \tb{d}_1^*   &      &  \iu \tb{t}_p' 
    \end{array}
    \right].
\end{align}
Here shaking induces transitions between two orbitals on two neighboring sites
with coupling strength
\begin{equation}
\tb{d}_1=\int d\mathbf{x}
\mathcal{W}^*_{s}(\mathbf{x})\hat p_x\mathcal{W}_{p_x}(\mathbf{r}+\hat x).
\end{equation}
Note that there are also diagonal terms given by
\begin{align}
\iu \tb{t}_s'&= \int d\mathbf{x}\mathcal{W}^*_{s}(\mathbf{x})\hat p_x
\mathcal{W}_{s}(\mathbf{r}+\hat x),\\ 
\iu \tb{t}_p'&= \int d\mathbf{x}
\mathcal{W}^*_{p_x}(\mathbf{x})\hat p_x\mathcal{W}_{p_x}(\mathbf{x}+\hat x). \label{itpp}
\end{align}
The matrix elements of $V(\tau)$ obtained here Eq.~(\ref{vtau}-\ref{itpp}) are
crucial for our subsequent analysis. 

\begin{figure}[h]
  \centering
  \includegraphics[width=0.45\textwidth]{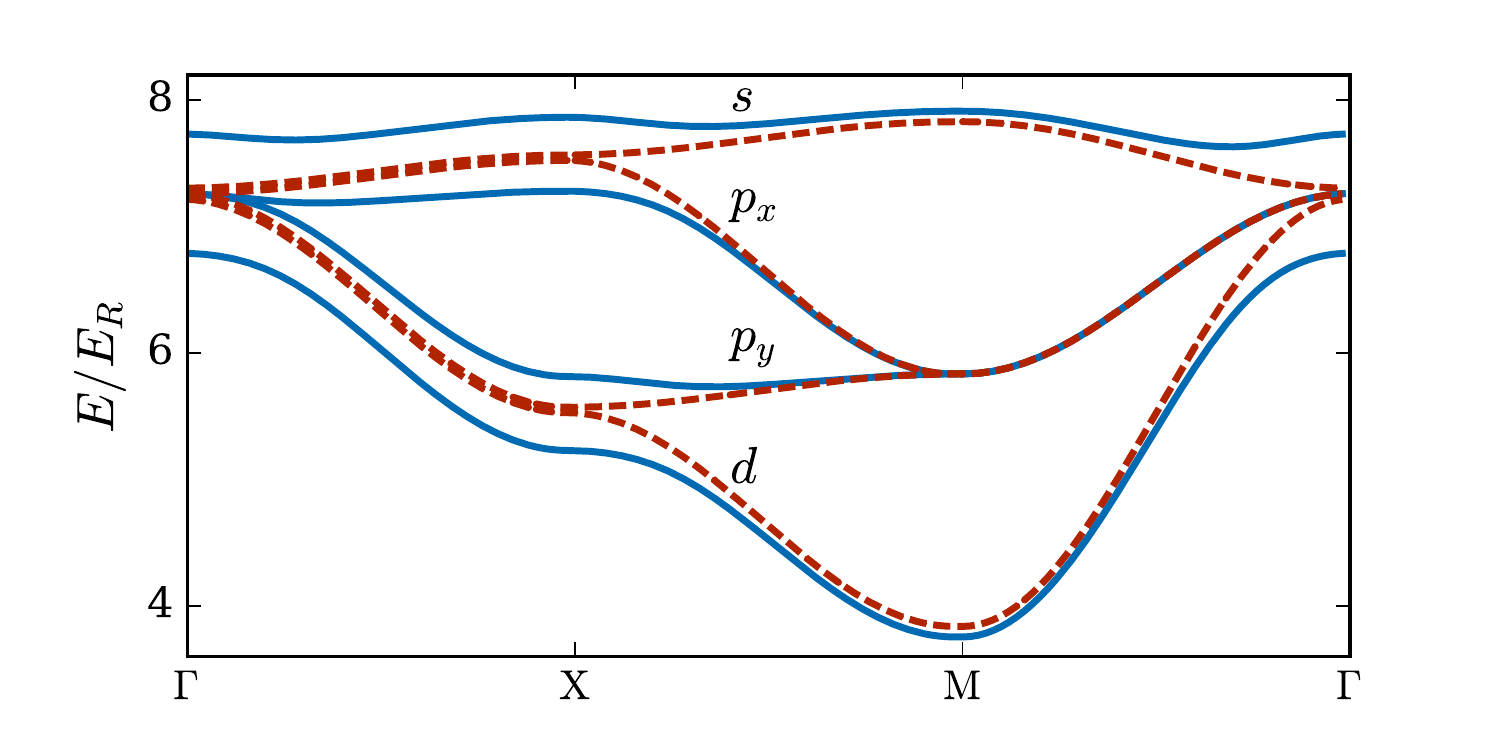}
  \includegraphics[width=0.45\textwidth]{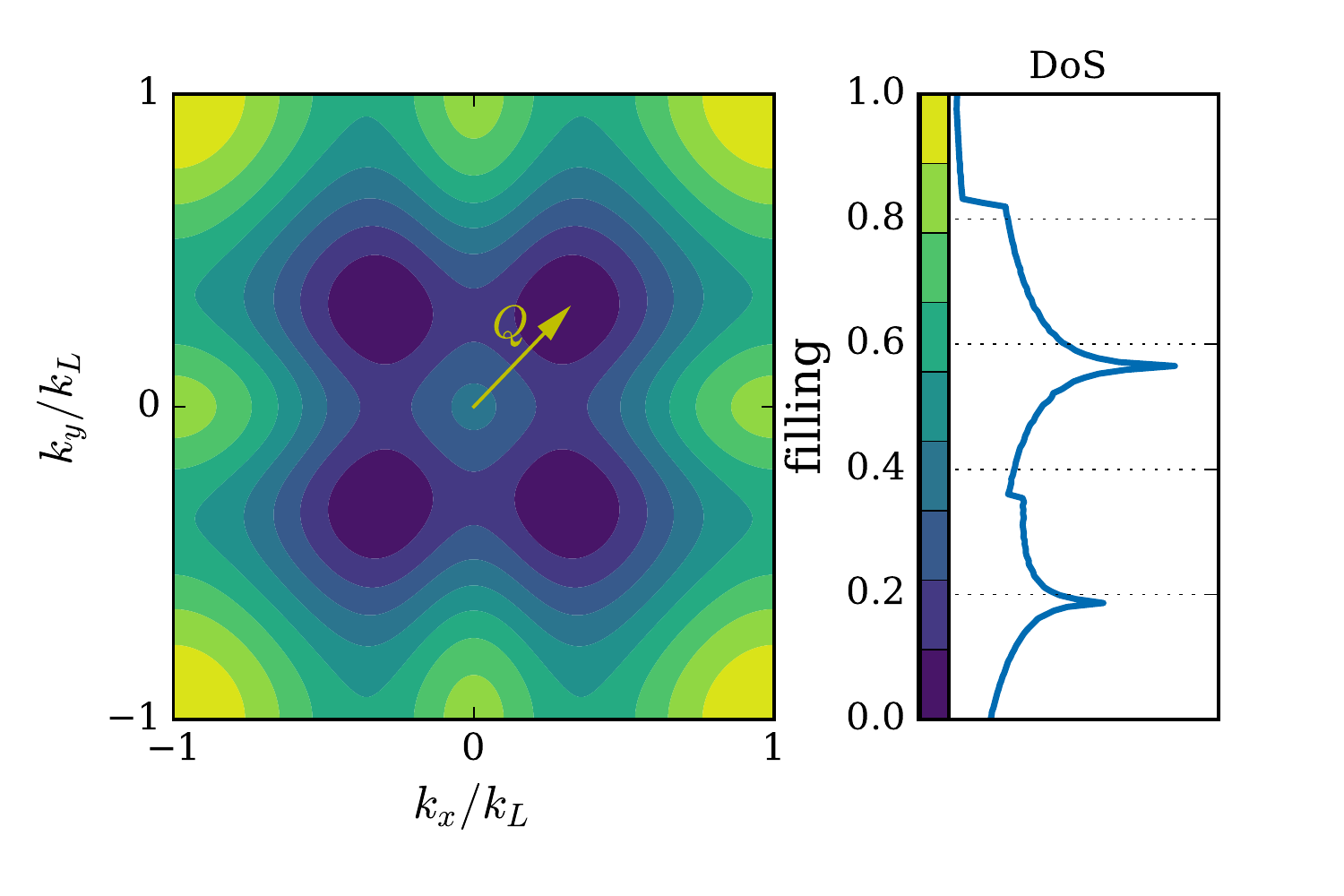}
  \includegraphics[width=0.45\textwidth]{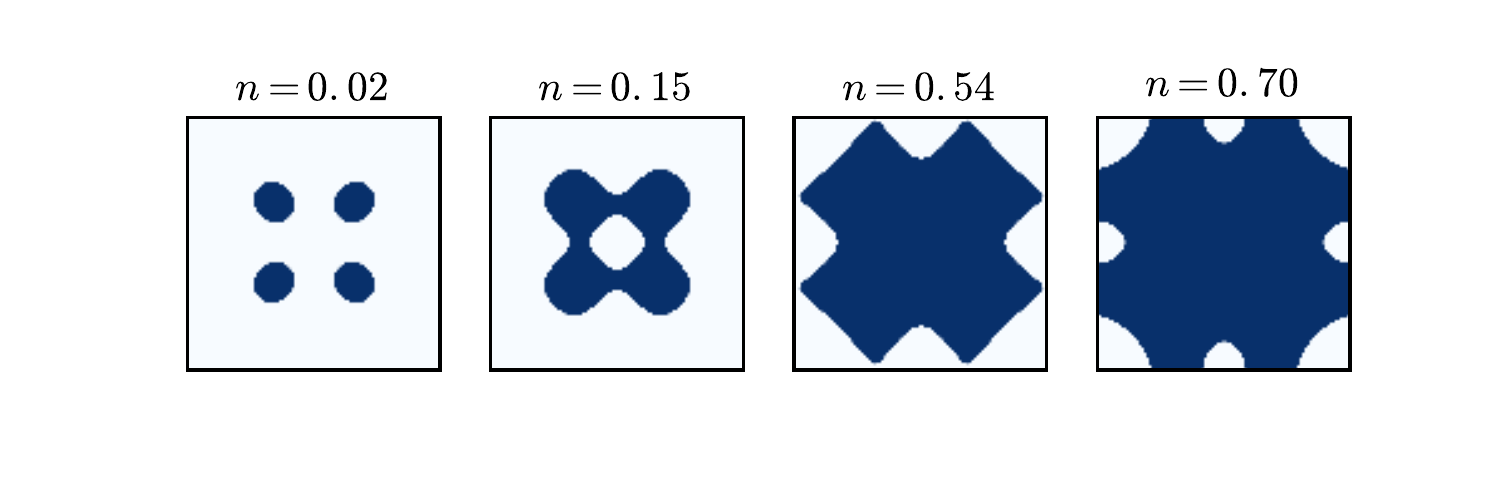}
  \caption{(Color online) Band structure of the effective four-band model
    $H_\mathrm{eff}$. TOP: Hybridization of four lowest Bloch bands by
    shaking.  The dashed (red) lines are the bare bands folded into same QeBZ,
    see Eq.~\eqref{bare_energies}, that will be coupled when shaking is turned
    on. The resulting hybridized
    energies,  $\epsilon_{\kappa\kappa'}(\kb)$ in Eq.
    \eqref{eq:effective_dispersion}, are shown by the solid (blue) lines.
    MIDDLE: Energy spectrum of the hybridized $s$-band in the first Brillouin
    zone (left) and the corresponding density of states at the Fermi level for
    a given lattice filling (right). BOTTOM: Evolution of Fermi surface with
    increasing lattice filling.  Parameters used are $e_s=-0.56, t_s= -0.07,
    e_p= 2.76, t_p=0.42, d_0=2.63, d_1=-0.28$ corresponding to $V_0/E_R=5$.
    Shaking frequency $\omega = 1.01\Omega$ whereas the amplitude is
    $\beta=0.1$.  }
  \label{fig:dispersion}
\end{figure}

To obtain a time-independent effective Hamiltonian, we use the RWA
$\psi_n\rightarrow[U_R]_{nn'}\psi_{i,n'}$ for $\psi$ in
Eq.~\eqref{eq:H_wannier_expansion} for a given site $i$.  The transformation
matrix is given by
\begin{equation}
U_R=\mathrm{Diag}(\ee^{\iu2\omega\tau}, \ee^{\iu\omega\tau},
\ee^{\iu\omega\tau}, 1), 
\end{equation}
where the choice of the exponentials in the matrix $U_R$ is motivated by
our numerical results in Fig.~\ref{fig:shaken_band}. Specifically, the band gap
between $s$ and $p$-bands is of the same order of the band gap between $p$ and
$d$-bands. Thus near resonance shaking couples the $s$ and $p$-bands, and also the $p$
and $d$-bands. $U_R$ accounts the interplay between these four orbitals by
shifting $s$-band by energy $2\omega$ and $p$-bands by $\omega$
such that all four levels are within the same energy window. The second step
of RWA is to drop remaining rapidly oscillating terms in $U^\dagger_R
H'_0 U_R$. In particular, we find the diagonal terms $\iu t_{s,p}'$ in
$V_{1}$ are removed by RWA. After Fourier transformation to momentum
space, the resulting effective Hamiltonian for the shaken system takes a clean
form,
\begin{align}
  H_\mathrm{eff} &= 
    \left[
    \begin{array}{cccc}
      E_s& D_x & D_y &  0   \\
       D_x & E_{p_x}& 0 & D_y \\
       D_y^* & 0 & E_{p_y}& D_x  \\
        0 & D_y^*  &  D_x^*  & E_d
    \end{array}
    \right].
    \label{eq:rwa_hamiltonian}
\end{align}
Here diagonal elements are the bare energies of four bands 
(see the dashed curves in Fig. \ref{fig:dispersion}) 
\begin{align}
E_s(\mathbf{k})&=\varepsilon_s(k_x)+\varepsilon_s(k_y)+2\omega, 
\nonumber\\
E_{p_x}(\mathbf{k})&=\varepsilon_p(k_x)+\varepsilon_s(k_y)+\omega,
\nonumber\\
E_{p_y}(\mathbf{k})&=\varepsilon_s(k_x)+\varepsilon_p(k_y)+\omega,
\nonumber\\
E_d(\mathbf{k})&=\varepsilon_p(k_x)+\varepsilon_p(k_y) \label{bare_energies}
\end{align}
where $\varepsilon_s(k_\mu) = \tb{e}_s
+ 2\tb{t}_s\cos(k_\mu)$ and $\varepsilon_p(k_\mu) = \tb{e}_p +
2\tb{t}_p\cos(k_\mu)$, $\mu=x,y$.
The off-diagonal terms are inter-orbital couplings
induced by shaking,
\begin{align}
D_{x}&=\beta\left[ d_0+2d_1\cos(k_{x})\right], \\
D_{y}&=\beta\ee^{\iu\vartheta} \left[ d_0+2d_1\cos(k_{y})\right].
\end{align} 
For given $V_0/E_R$, we can calculate parameters $e_n$, $t_{s,p}$,
$d_{0,1}$ numerically from the Wannier functions constructed from the Bloch 
waves. For reference we provided a few typical values of these 
parameters in Table~\ref{tab:table1}.

\begin{table}[t!]
  \setlength{\tabcolsep}{0.8em}
  \centering
  \caption{Numerical values of the parameters in the effective Hamiltonian $
    H_\mathrm{eff}$ for shaken square lattice 
    for different lattice depth $V_0$ in units of recoil
    energy $E_R$. $e_n$ and $t_n$ are onsite energies and nearest neighbor
    hopping consistent with Ref.~\cite{Li2015}. $d_0$ and $d_1$ are the onsite and
    nearest neighbor inter-orbital coupling strength, respectively. }
  \label{tab:table1}
  \begin{tabular}{ccccccc}
    \hline
    $V_0$  & $e_s$ & $e_p$ & $t_s$ & $t_p$ & $d_0$ & $d_1$ \\
    \hline
    3  & -0.06 & 2.57 & -0.11 & 0.50 & 2.07 & -0.44 \\
    5  & -0.56 & 2.76 & -0.07 & 0.42 & 2.63 & -0.28 \\
    10 & -2.12 & 2.94 & -0.02 & 0.24 & 3.47 & -0.07 \\
    20 & -5.80 & 1.98 & -0.00 & 0.06 & 4.37 & -0.01 \\ 
    \hline
  \end{tabular}
\end{table}

Eq. \eqref{eq:rwa_hamiltonian} is one of the main results of this paper. We
can further rewrite $H_\mathrm{eff}$ as the sum of two direct products of the
form
\begin{equation}
    H_\mathrm{eff} = H_x\otimes \unity_y + \unity_x \otimes H_y
\end{equation}
with $\unity_\mu$ the two-by-two unit matrix and $H_\mu$ given by
\begin{align}
    H_x = 
  \left[
    \begin{array}{cc}
      \zeta_x + h_x   &  D_y  \\
      D_y^*   & \zeta_x - h_x 
    \end{array}
  \right],\;\;
    H_y = 
  \left[
    \begin{array}{cc}
      \zeta_y + h_y   &  D_x  \\
      D_x^*   & \zeta_y - h_y 
    \end{array}
  \right],
\end{align}
where $\zeta_\mu = [\varepsilon_s(k_\mu)+\omega+\varepsilon_p(k_\mu)]/2$ and
$h_\mu=[\varepsilon_s(k_\mu)+ \omega - \varepsilon_p(k_\mu)]/2$.
$H_\mathrm{eff}$ is diagonalized by a
unitary transformation
\begin{equation}
  \mathcal{U}(\kb)=\mathcal{U}_x(\kb)\otimes\mathcal{U}_y(\kb),
  \label{eq:diagonalization}
\end{equation}
where
\begin{align}
  \mathcal{U}_\mu(\kb) = 
  \left[
    \begin{array}{cc}
      \cos\theta_\mu\ee^{\iu\varphi_\mu} &  
      -\sin\theta_\mu\ee^{\iu\varphi_\mu}\\
      \sin\theta_\mu & \cos\theta_\mu
    \end{array}
  \right].
\end{align}
The two angles here, $\theta_\mu$ and $\varphi_\mu$, are defined by
\begin{align}
  \cos(2\theta_{x,y})&=\frac{h_{x,y}}{\sqrt{h_{x,y}^2+|D_{y,x}|^2}}, \label{thetamu}\\
%  \sin(2\theta_{x,y})&=\frac{|D_{y,x}|}{\sqrt{h_{x,y}^2+|D_{y,x}|^2}},\\
\ee^{\iu\varphi_{x,y}} &= \frac{D_{y,x}}{|D_{y,x}|}.
\end{align}
Notice that $D_y$ couples to $h_x$ and $D_x$ couples to $h_y$ in the definitions of $\theta_\mu$ and
$\varphi_\mu$. The four eigenvalues of $H_\mathrm{eff}$ are found as
\begin{equation}
  \epsilon_{\kappa\kappa'}(\kb) = \zeta_x + \zeta_y + 
  \kappa\sqrt{h^2_x+|D_y|^2}+
  \kappa'\sqrt{h^2_y+|D_x|^2}
  \label{eq:effective_dispersion}
\end{equation}
where $\kappa,\kappa'=\pm$. For example, the hybridized $s$-band, modified from
the bare $s$-band due to mixing with other bands by shaking, is given by
\begin{equation}
 \epsilon_1(\kb)\equiv \epsilon_{++}(\kb).
\end{equation}
Let us introduce the fermion creation operators in the basis of
$H_\mathrm{eff}$, $\Phi\dgr=(\phi\dgr_1,\phi\dgr_2,\phi\dgr_3, \phi\dgr_4)$,
with $\phi\dgr_1$ corresponding to the hybridized $s$-band $\epsilon_1(\kb)$
for example (the momentum and spin index are suppressed for brevity). They are
related to the creation operators in the original basis
$\Psi\dgr\equiv(\psi_{s}\dgr,\psi_{p_x}\dgr,\psi_{p_y}\dgr,\psi_{d_{xy}}\dgr)$
by the unitary transformation $\Psi_{ \kb} = \mathcal{U}(\kb)\Phi_{\kb}$ or
$\psi_{n\kb}=\mathcal{U}_{nm}(\kb)\phi_{ m \kb}$ where $\mathcal{U}(\kb)$ is
given in Eq.~\eqref{eq:diagonalization}. This relation will become important
in the next section.

The band structure of $H_\mathrm{eff}$ is illustrated in
Fig.~\ref{fig:dispersion} for lattice depth $V_0/E_R=5$. The top row compares
the four energy bands (solid lines) described by
$\epsilon_{\kappa\kappa'}(\kb)$ in Eq.~\eqref{eq:effective_dispersion} with
the bare band dispersions (dashed lines) given in Eq.~\eqref{bare_energies}.
One can see that the level repulsion between the $s$-, $p$- and $d$-bands
pushes the $s$-band up around the $\Gamma$ point. The dispersion of the
hybridized $s$-band in the entire 2D Brillouin zone and the corresponding
density of states are shown in the middle row of Fig.~\ref{fig:dispersion}.
The four band minima are clearly seen here. The spectrum obtained here is in
good agreement with the numerical solution in section \ref{sec:numerics}.
When the hybridized $s$-band is gradually filled with fermions, the
resulting Fermi surfaces undergoes a non-trivial evolution as shown in the
bottom row of Fig.~\ref{fig:dispersion}. The Fermi surface topology change
found here highlights the capability of lattice shaking in engineering the
band structures. 

The analytical form of the effective Hamiltonian $H_\mathrm{eff}$ and the
resulting spectrum clarify the physics of shaken square optical lattice. It
captures succinctly how the relevant orbitals, either on the same site or two
neighboring sites, are coupled by shaking. The simplicity achieved is partly
due to our choice of a convenient gauge, where $H'_0$ splits into
$H_\mathrm{static}$ and
$V(\tau)=\dot{\mathbf{x}}_0(\tau) \cdot\mathbf{p}$. It is also derived from
the symmetries of the Wannier functions and $V(\tau)$, leading to for example
only two independent coupling strength $d_0$ and $d_1$.  These results from
the treatment of circularly shaken square lattice may be useful for the study
of other lattice geometries and shaking protocols.

%}}}
\section{Effective Interactions }%{{{
\label{sec:interactions}

The peculiar Fermi surfaces found in the previous section suggests that
interaction may drive interesting many-body instabilities for Fermi gases in
shaken lattices. The effective interactions for two fermions on the Fermi
surface of the hybridized bands will differ from the bare interactions.  In
this section, we outline a procedure to derive these effective interactions
$V_\mathrm{eff}$ and then work out its explicit expression for fermions on the
hybridized $s$-band.  We only consider weak interactions of spin-$1/2$
fermions in the sense that the energy scale related to interactions is assumed to be
much smaller than the shaking frequency and the band-width. In other words, we
will treat interaction as a weak perturbation to $H_\mathrm{eff}$ in
Eq.~\eqref{eq:rwa_hamiltonian}.

We first expand the contact interaction in Eq.~\eqref{eq:interaction}
in the Wannier basis using Eq. \eqref{eq:wannier_expansion}. It then takes
the Hubbard-like onsite form in the multi-orbital basis,
\begin{equation}
  V_I = \sum_i\sum_{nmm'n'}U_{nmm'n'}
  \psi_{\uar n i}\dgr %(j)
  \psi_{\dar m i}\dgr %(j)
  \psi_{\dar m'i}     %(j)
  \psi_{\uar n'i}     %(j)
\end{equation}
where the time dependence of $\psi$ is suppressed for brevity and $U_{nmm'n'}
\equiv g\int d\x \W_n^*(\x) \W_m^*(\x) \W_{m'}(\x) \W_{n'}(\x)$.  In
accordance with the previous section, we only keep the orbitals
$n=s,p_x,p_y,d_{xy}$. Also we will approximate the Wannier functions with
local harmonic oscillator eigenstates to evaluate the integrals $U_{nmm'n'}$.
This is justified for deep lattices and it simplifies the algebra greatly. In
fact, as we will show below, all the onsite interactions in different orbital
channels can be expressed in terms of the $s$-orbital interaction constant
$$U\equiv g\int d\x |\W_s(\x)|^2 |\W_{s}(\x)|^2.$$

Next we address the question what happens to the interaction term $V_I$ during
the RWA. To answer this question in a transparent way,
we split the terms in $V_I$ into three distinct channels
\begin{equation}
    V_I = U\big(V_\mathrm{density} + 
    V_\mathrm{ex} + V_\mathrm{pt}\big).\label{triple}
\end{equation}
The first term is the density-density interaction given by
%\begin{align}
%  V_\mathrm{density} &= 
%  \mathrm{n}_{\uar s}\mathrm{n}_{\dar s}+
%  \frac{3}{4}
%  \big[
%  \mathrm{n}_{\uar x}\mathrm{n}_{\dar x} + \mathrm{n}_{\uar x}\mathrm{n}_{\dar x}
%  \big]+
%  \frac{9}{16}
%  \mathrm{n}_{\uar d}\mathrm{n}_{\dar d} \nonumber\\
%  &+\frac{1}{4}\mathrm{n}_s(\mathrm{n}_x+\mathrm{n}_y) 
%  +\frac{1}{8}\big[\mathrm{n}_s\mathrm{n}_d + 
%  \mathrm{n}_x\mathrm{n}_y\big] \nonumber\\
%  &+\frac{3}{16}(\mathrm{n}_x+\mathrm{n}_y)\mathrm{n}_d.
%\end{align}
%or
\begin{align}
  V_\mathrm{density} = 
  \vec{\n}_\uar\T
  \cdot \hat G_\mathrm{density} \cdot\vec{\n}_\dar
\end{align}
where $\vec{\n}_\sigma\T=[\n_{\sigma s}, \n_{\sigma x}, n_{\sigma y},
\n_{\sigma d}]$, the superscript $\top$ denotes matrix transposition and the
density operator is defined as $\mathrm{n}_{\sigma n} = \psi_{\sigma
  n}^\dagger\psi_{\sigma n}$ with the site index $i$ dropped for brevity.  The
elements of the constant matrix
%we defined
\begin{equation}
  \hat G_\mathrm{density} =%\frac{1}{16}
%  \left[
%    \begin{array}{cc}
%      4   &  2  \\
%      2   & 3 
%    \end{array}
%  \right]\otimes
%  \left[
%    \begin{array}{cc}
%      4   &  2  \\
%      2   & 3 
%    \end{array}
%  \right].
  % or
  \left[
    \begin{array}{cccc}
      1   &   1/2   &   1/2   &  1/4  \\
      1/2   &   3/4   &   1/4   &  3/8  \\
      1/2   &   1/4   &   3/4   &  3/8  \\
      1/4   &   3/8   &   3/8   & 9/16 
    \end{array}
  \right]
  \label{eq:Gdensity}
\end{equation}
%a=[\mathrm{n}_{\sigma s},\mathrm{n}_{\sigma x},\mathrm{n}_{\sigma y},\mathrm{n}_{\sigma d}]$
%, $\mathrm{n}_n=\mathrm{n}_{\uar n}+\mathrm{n}_{\dar n}$. 
are obtained by evaluating the overlap integrals using the approximate Wannier
functions. Now consider the effect of RWA, $[\psi_{\sigma s}, \psi_{\sigma x},
\psi_{\sigma y}, \psi_{\sigma d}]\overset{\text{\tiny RWA}}{\rightarrow}
[\ee^{-2\iu\omega\tau}\psi_{\sigma s}, \ee^{-\iu\omega\tau}\psi_{\sigma x},
\ee^{-\iu\omega\tau}\psi_{\sigma y}, \psi_{\sigma d}]$, on these terms. One can see that for
every $\psi_{\sigma n}$ operator, there is a corresponding $\psi_{\sigma
  n}\dgr$. Therefore no time dependent terms 
like $\ee^{-\iu\omega\tau}$ and $\ee^{-2\iu\omega\tau}$ will remain, 
thus $V_\mathrm{density}$ is invariant under RWA.

The second term is the orbital exchange interaction
%\begin{align}
%  V_\mathrm{spin} &= 
%   -\mathbf{S}_s\cdot\mathbf{S}_x -\mathbf{S}_s\cdot\mathbf{S}_y
%   -\frac{1}{2}\mathbf{S}_s\cdot\mathbf{S}_d
%   -\frac{1}{2}\mathbf{S}_x\cdot\mathbf{S}_y \nonumber\\
%   &-\frac{3}{4}\mathbf{S}_x\cdot\mathbf{S}_d
%   -\frac{3}{4}\mathbf{S}_y\cdot\mathbf{S}_d
%\end{align}
% or
\begin{align}
  V_\mathrm{ex} = -
  \vec{\s}^+ \cdot \hat G_\mathrm{ex} \cdot\vec{\s}^-
\end{align}
where 
\begin{equation}
  \hat G_\mathrm{ex} =%\frac{1}{16}
%  \left[
%    \begin{array}{cc}
%      4   &  2  \\
%      2   & 3 
%    \end{array}
%  \right]\otimes
%  \left[
%    \begin{array}{cc}
%      4   &  2  \\
%      2   & 3 
%    \end{array}
%  \right].
  % or
  \left[
    \begin{array}{cccc}
      0   &   1/2   &   1/2   &  1/4  \\
      1/2   &   0   &   1/4   &  3/8  \\
      1/2   &   1/4   &   0   &  3/8  \\
      1/4   &   3/8   &   3/8   & 0 
    \end{array}
  \right], 
  \label{eq:Gex}
\end{equation}
with $\vec{\s}^+=[\s^+_s, \s^+_x, \s^+_y, \s^+_d]$, and $\vec{\s}^-=[\s^-_s,
\s^-_x, \s^-_y, \s^-_d]\T$. The raising and lowering operators are defined as
usual
$\s_n^\pm = \s_n^1\pm\iu\s_n^2$, with $\s_n^\mu=\frac{1}{2}
\gamma^\mu_{\sigma\sigma'}\psi_{\sigma n}\dgr\psi_{\sigma' n}$ and
$\boldsymbol{\gamma}=(\gamma^1,\gamma^2,\gamma^3)$ are the Pauli matrices.
Similar to density-density interactions, one can see that
there is a $\psi_{\sigma
  n}\dgr$ for every $\psi_{\sigma' n}$ operator.  Thus, $V_\mathrm{ex}$ is
also invariant under RWA.

Finally, the last term in Eq. \eqref{triple} describes pair transfers between
the orbitals,
\begin{align}
  V_\mathrm{pt} =
  \vec{\p}\dgr \cdot \hat G_\mathrm{pt} \cdot\vec{\p},
\end{align}
where $\vec{\p}\dgr=[\p_s\dgr, \p_x\dgr, \p_y\dgr, \p_d\dgr]$,  $\p_n\dgr =
\psi_{\uar n}\dgr\psi_{\dar n}\dgr$ is the pair creation operator for orbital
$n$, and $\hat G_\mathrm{pt}$ is identical to $\hat G_\mathrm{ex}$ above.
The pair transfer between the $p_x$ and $p_y$ orbitals is invariant under RWA
since they have the same exponential time dependence.  However, for pair
transfers between $s$ and $p$, $s$ and $d$ as well as $p$ and $d$ orbitals,
time dependent exponentials will remain after the RWA.  Since such fast
oscillating terms are subsequently ignored in RWA, $\hat G_\mathrm{pt}$
becomes simplified,
\begin{equation}
    \hat G_\mathrm{pt}\overset{\text{\tiny RWA}}{\rightarrow}
  \left[
    \begin{array}{cccc}
      0   &   0   &   0   &  0  \\
      0   &   0   &   1/4   &  0  \\
      0   &   1/4   &   0   &  0  \\
      0   &   0   &   0   & 0 
    \end{array}
  \right]. \label{Gpt}
\end{equation}

It follows that after the RWA, the (time-independent)
effective interaction in the momentum space takes the following form,
\begin{equation}
  V'_I = 
  \sum_{\substack{ \kb\kb'\\
                   nmm'n' }}
  U'_{nmm'n'}
  \psi_{\uar n  \kb }\dgr
  \psi_{\dar m -\kb }\dgr
  \psi_{\dar m'-\kb'}
  \psi_{\uar n' \kb'}.
\end{equation}
Here $\psi$ no longer depends on $\tau$, and $U'_{nmm'n'}$ differs from
$U_{nmm'n'}$ by absence of all pair transfer terms but the one in between
$p_x$ and $p_y$ orbitals, in accordance with Eq. \eqref{Gpt}. $U'$ can be
straightforwardly constructed from the $\hat{G}$ matrices above
[Eqs.~\eqref{eq:Gdensity}, \eqref{eq:Gex} and \eqref{Gpt}] and its various
terms will not be tabulated here.

The last step is to rewrite $V'_I$ in terms of the field operators in the
basis of $H_\mathrm{eff}$. This is achieved by the unitary transformation,
$\psi_{\sigma n\kb}=\mathcal{U}_{nm}(\kb)\phi_{\sigma m \kb}$, with
$\mathcal{U}(\kb)$ given in Eq.~\eqref{eq:diagonalization}.  As an example,
let us focus on the effective interactions for two fermions of
opposite momenta on the hybridized $s$-band, denoted with $V^s_I$  below. For this
purpose, we can
project $V'_I$ onto the $n=1$ band by substituting $\psi_{\sigma m
  \kb}=\mathcal{U}_{m1}\phi_{\sigma 1\kb}$ and $\psi\dgr_{\sigma m\kb} =
\phi\dgr_{\sigma 1\kb}\mathcal{U}\dgr_{1m\kb}$ into the expression for $V'_I$
and collecting the relevant terms to get
\begin{equation}
  V^s_I = U\sum_{\kb\kb'}
  V_\mathrm{eff}(\kb,\kb')
  \phi_{\uar1 \kb }\dgr
  \phi_{\dar1-\kb }\dgr
  \phi_{\dar1-\kb'}
  \phi_{\uar1 \kb'}.
\end{equation}
where we have factored out onsite $s$-band interaction constant $U$ for
convenience and defined the interaction vertex $V_\mathrm{eff}(\kb,\kb')$ as
\begin{align}
  V_\mathrm{eff}(\kb,\kb') = 
    \sum_{nmm'n'}
  &U'_{nmm'n'}
  \mathcal{U}^*_{n 1}( \kb )
  \mathcal{U}^*_{m 1}(-\kb )
  \nonumber\\
  &\times
  \mathcal{U}  _{m'1}(-\kb')
  \mathcal{U}  _{n'1}( \kb').
  \label{eq:effective_interaction}
\end{align}
This expression can be simplified using the matrix elements of $\mathcal{U}(\kb)$ 
in Eq.~\eqref{eq:diagonalization} with some algebra as
\begin{equation}
    V_\mathrm{eff}(\kb,\kb') = 
    V^s_\mathrm{density}(\kb,\kb')+
    V^s_\mathrm{ex}(\kb,\kb')+
    V^s_\mathrm{pt}(\kb,\kb').
    \label{eq:full_interaction}
\end{equation}
The first two terms are given by
\begin{align}
    V^s_\mathrm{density}(\kb,\kb') &=(A_x+B_x)(A_y+B_y),\\
    V^s_\mathrm{ex}(\kb,\kb') &=A_xB_y+B_xA_y+B_xB_y,
\end{align}
in terms of
\begin{align}
    A_\mu &=\cos^2\theta_\mu\cos^2\theta_\mu'+\frac{3}{4}
    \sin^2\theta_\mu\sin^2\theta_\mu', \\
    B_\mu &=\frac{1}{4} \sin2\theta_\mu\sin2\theta_\mu'.
    \end{align}
where $\theta_\mu$ is defined in Eq.~\eqref{thetamu}.  
And the pair transfer term takes
the following form,
\begin{equation}
    V^s_\mathrm{pt}(\kb,\kb') =
    \frac{1}{4}
    \left[
      \cos\theta_x'\sin\theta_y'\sin\theta_x\cos\theta_y
    \right]^2+
      x\leftrightarrow y.
\end{equation}
Note that we have considered the interactions between two particles with zero
center of mass momentum above. It is straightforward to obtain more general
interaction vertices of the form $$V(\kb_1,\kb_2,\kb_3)
  \phi_{\uar1, \kb_1+\kb_2-\kb_3 }\dgr
  \phi_{\dar1, \kb_3}\dgr
  \phi_{\dar1, \kb_2}
  \phi_{\uar1, \kb_1}$$ by a similar projection procedure. The result is rather
  lengthy and will not be given here.
  
\begin{figure}[h]
  \centering
  \includegraphics[width=0.22\textwidth]{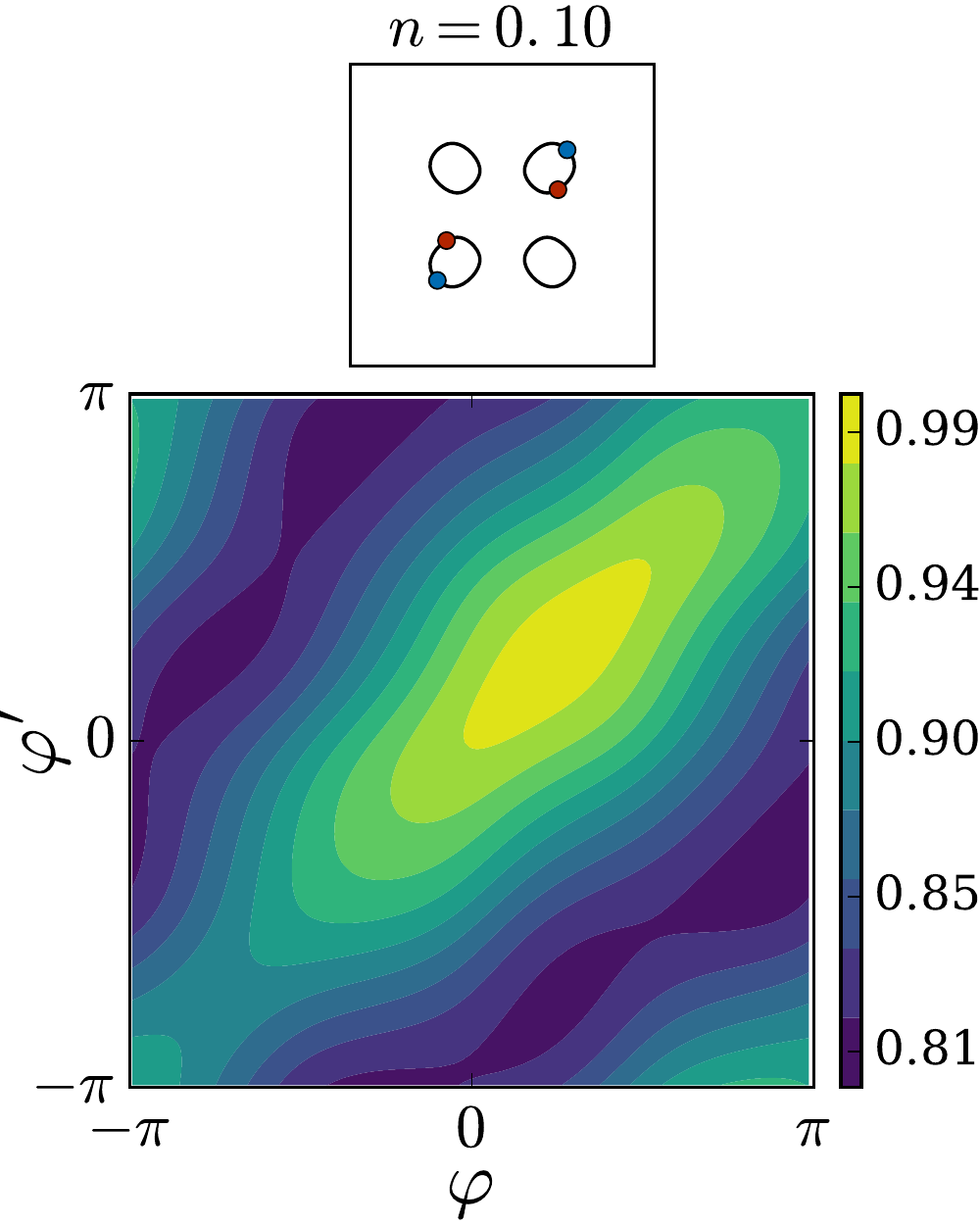}
  \includegraphics[width=0.22\textwidth]{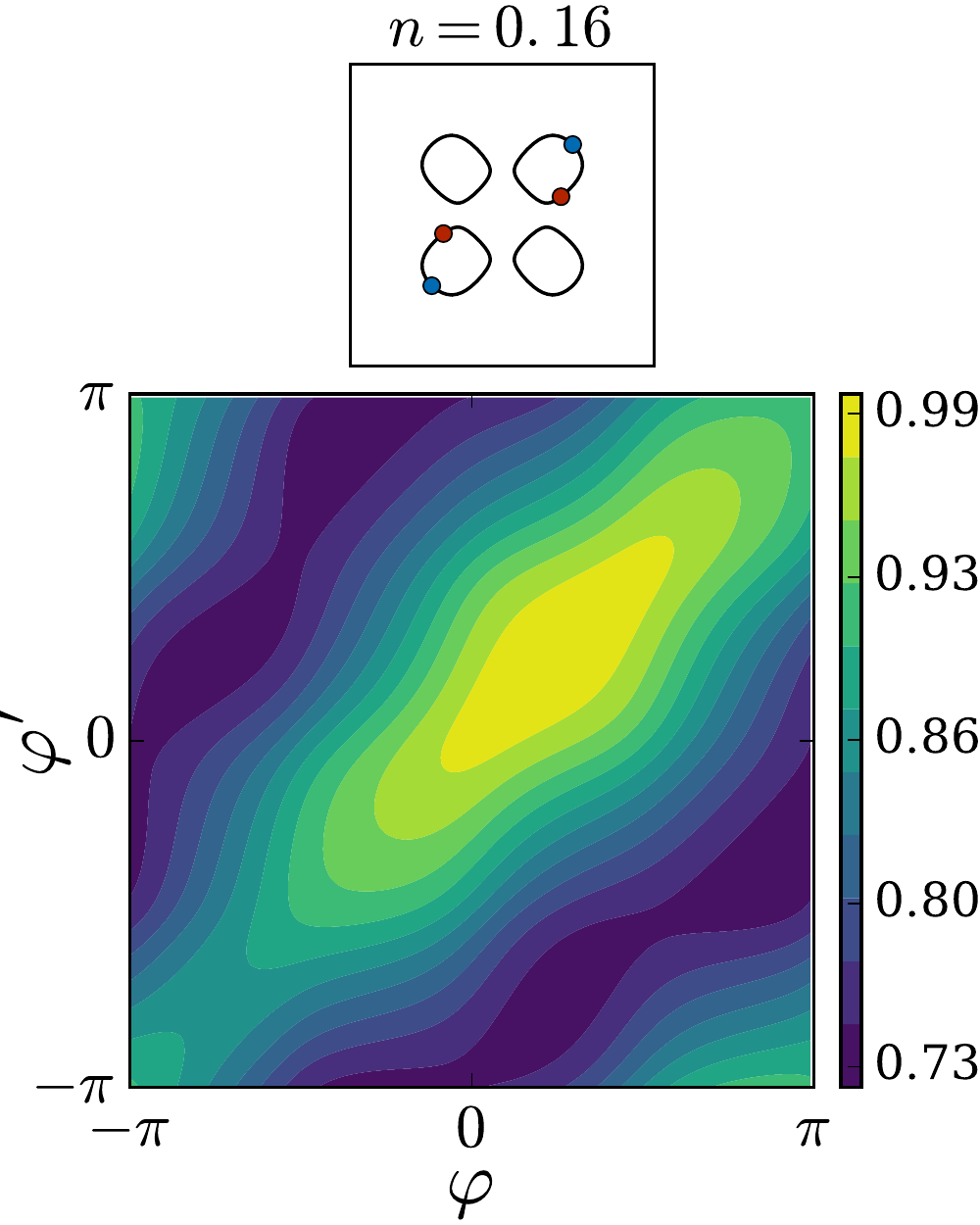}
  \includegraphics[width=0.22\textwidth]{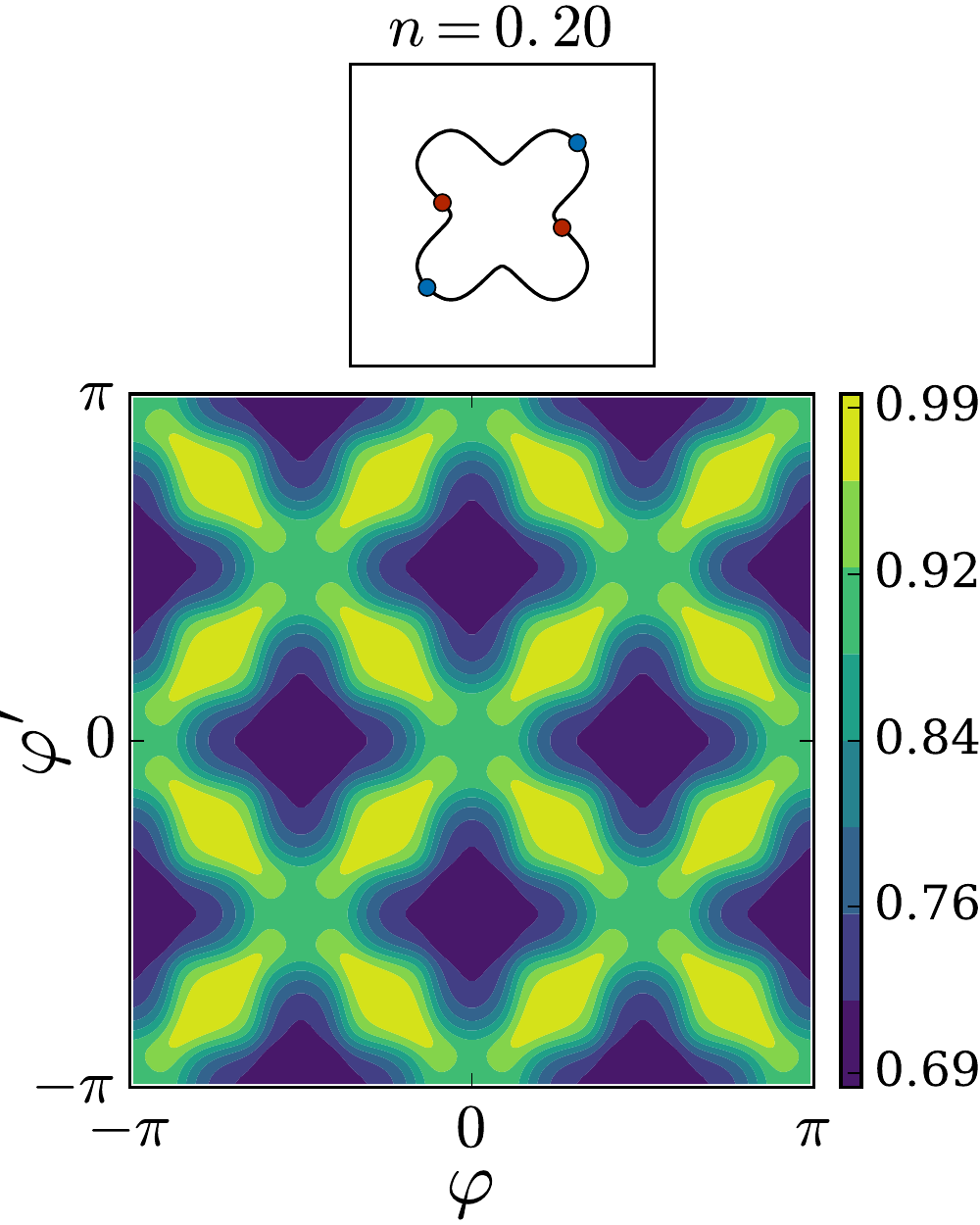}
  \includegraphics[width=0.22\textwidth]{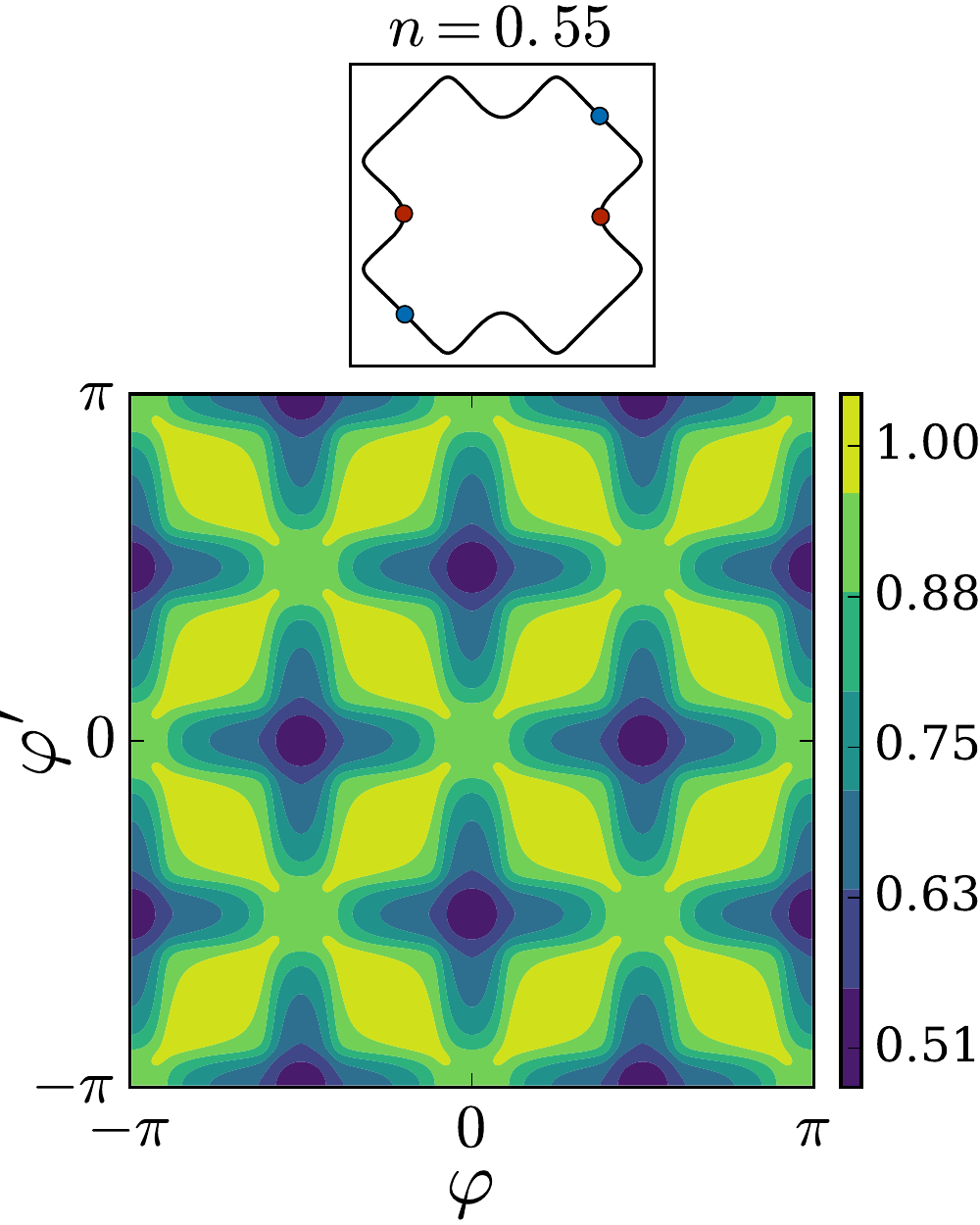}
  \caption{(Color online) Momentum dependence of the effective interaction on
    the Fermi surface $V_\mathrm{eff}(\kb,\kb')\equiv V(\varphi,\varphi')$ for
    four different fillings $n$. The shaking parameters are $V_0=5E_R$,
    $\beta=0.1$, $\omega=1.01\Omega$.  Here $\varphi$ and $\varphi'$ are polar
    angles of $\kb$ and $\kb'$ on the Fermi surface (solid lines)
    respectively, defined with respect to the center of the entire Fermi
    surface (lower panel) or the Fermi pocket in the first quadrant (upper
    panel).  The red and blue dots depict examples for the definitions of
    angles corresponding to momentum pair $(\kb,-\kb)$ for minimum and maximum
    interaction angles, respectively. Note that the angles are defined from
    $-\pi$ to $\pi$.
    }
  \label{fig:interaction}
\end{figure}

The most interesting property of $V_\mathrm{eff}(\kb,\kb')$ is its nontrivial
$\mathbf{k}$ dependence.  This is in marked contrast with the bare interaction
which is constant in $\mathbf{k}$ space.  In analogy with the hybridized band
dispersion discussed in the previous section, we may say that the interaction
is strongly modified by shaking induced band mixing.
Fig.~\ref{fig:interaction} shows a few examples of $V_\mathrm{eff}(\kb,\kb')$
for different fillings, where $\kb$ and $\kb'$ reside on the corresponding
Fermi surfaces.  For $n=0.55$, the variation in $V_\mathrm{eff}(\kb,\kb')$
reaches the order of 50\%.  Formally, the momentum dependence of
$V_\mathrm{eff}(\kb,\kb')$ is inherited from the $\kb$-dependence of the
$\mathcal{U}$ matrix under the projection procedure.  We can understand the
variation of $V_\mathrm{eff}(\kb,\kb')$ qualitatively as follows.  The
$p$-orbitals hybridize with the $s$-orbital in a non-uniform way around the
Fermi surface. In particular, the mixing is stronger near the  $\Gamma
\mathrm{X}$ line, where the ridge of the $p$-band come close to the $s$-band
(see Fig.~\ref{fig:dispersion}), than that along the $\Gamma \mathrm{M}$ line.
Since the onsite interaction constant of the bare $p$-bands is smaller than
the $s$-band due to the reduction in the overlap integrals, regions on Fermi
surface with more mixture of $p$-orbitals have smaller effective interaction.
Thus, we expect the effective interaction reaches maximum around, e.g.,
$\varphi_\kb\sim \varphi_{\kb'}\sim \pi/4$ (where $\varphi_\kb$ is the polar
angle on the Fermi surface) along the $\Gamma \mathrm{M}$ line, consistent
with the numerical results in Fig. \ref{fig:interaction}.

%}}}
\section{Pairing of Fermions in shaken square lattice}%{{{
\label{sec:pairing}

To summarize, we have arrived at the following effective Hamiltonian
for weakly interacting fermions in the hybridized $s$-band of the shaken lattice,
\begin{align}
  H_\mathrm{BCS} &= \sum_\kb \xi_\kb\phi\dgr_{\sigma\kb}\phi_{\sigma\kb}
  \nonumber\\
  &+\frac{1}{2} \sum_{\kb',\kb}V_\mathrm{eff}(\kb,\kb')
  \phi\dgr_{\uar\kb'}
  \phi\dgr_{\dar-\kb'}
  \phi_{\dar-\kb}
  \phi_{\uar+\kb}.
  \label{eq:effective_interacting_model}
\end{align}
Here we have dropped the orbital index $n=1$, $\phi_{\sigma
  1\kb}\rightarrow\phi_{\sigma\kb}$, and defined $\xi_\kb=\epsilon_1(\kb)-\mu$
  with chemical potential $\mu$. The band dispersion $\epsilon_1(\kb)$ is
  given in Eq.~\eqref{eq:effective_dispersion} and the reduced effective
  interaction $V_\mathrm{eff}(\kb,\kb')$ is given in
  Eq.~\eqref{eq:full_interaction}. 
We only consider weak, attractive interactions with $U<0$. The dominant
instability is Cooper pairing, and it is justified to only include
$V_\mathrm{eff}(\kb,\kb')$, i.e., the scattering of fermions with opposite
spins and
opposite momenta.  For strong coupling or repulsive interactions, one needs to
consider more general interaction vertices $V(\kb_1,\kb_2,\kb_3)$ and other
relevant instabilities.

Following the standard BCS theory, we define the pairing order parameter 
\begin{equation}
    \Delta(\kb) = -\sum_{\kb'} V_\mathrm{eff}(\kb,\kb')
    \langle 
  \phi_{\dar-\kb'}
  \phi_{\uar+\kb'} \rangle,
  \label{eq:gap}
\end{equation}
and perform mean-field decoupling of the quartic interaction terms in
Eq.~\eqref{eq:effective_interacting_model}. The resulting quadratic
Hamiltonian is diagonalized using Bogoliubov transformation, and the gap
equation becomes
\begin{equation}
    \Delta(\kb) = -\sum_{\kb'} V_\mathrm{eff}(\kb,\kb')
    \frac{\Delta(\kb')}{2E_{\kb'}}
    \tanh(E_{\kb'}/2k_BT)
\end{equation}
where $E_\kb=\sqrt{\xi_\kb^2+|\Delta(\kb)|^2}$ is the quasiparticle excitation
spectrum.  For temperatures close to the critical temperature $T_c$, the
magnitude of the gap is small. The gap equation can be linearized to become an
eigenvalue problem
\begin{equation}
    \int \frac{d\varphi_\kb'}{2\pi} V_\mathrm{eff}(\kb,\kb') \Delta(\kb') = 
    - \lambda\Delta(\kb).
    \label{eq:eigenvalue_equation}
\end{equation}
Here $\varphi_{\kb'}$ is the angle of the momenta $\kb'$ with respect to the
center of the Fermi surface. The largest eigenvalue $\lambda$ yields $T_c$
through the non-linear equation
\begin{equation}
    \frac{1}{\lambda} =  \frac{|U|}{2N_\mathrm{sites}}\sum_\kb 
    \frac{1}{\xi_\kb}\tanh(\xi_\kb/2k_BT_c). 
\end{equation}
And the corresponding eigenvector gives the orbital symmetry of the pairing
order parameter.  Here 
%$U$ is the onsite interaction constant and
$N_\mathrm{sites}$ is the number of lattice sites. For cases with multiple
Fermi surfaces (see bottom row of Fig.~\ref{fig:dispersion} for $n=0.15$ for
example), Eq.~\eqref{eq:eigenvalue_equation} is solved separately for
each Fermi surface such that the Fermi surface with
the largest eigenvalue determines the leading unstable surface. 

\begin{figure}[h]
  \centering
  \includegraphics[width=0.45\textwidth]{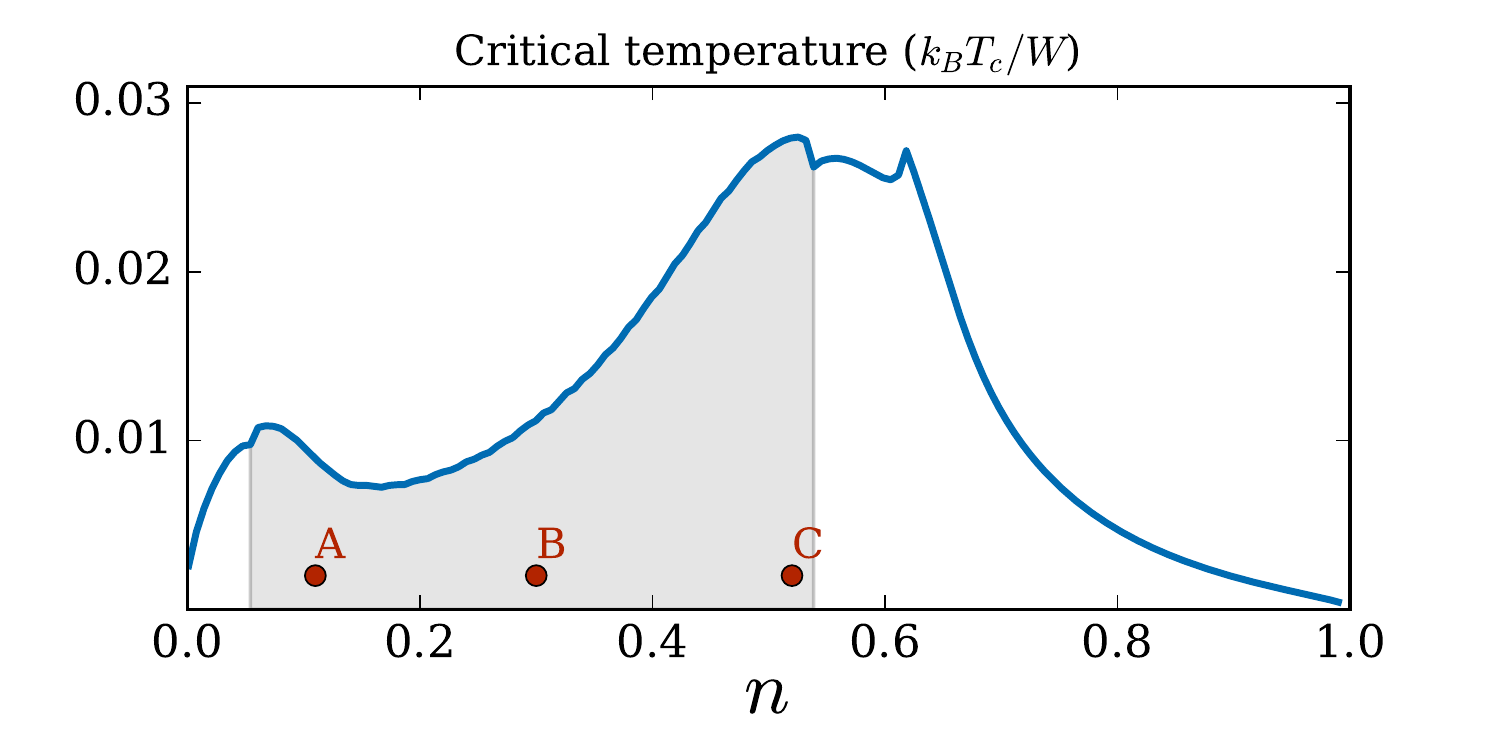}
  \includegraphics[width=0.45\textwidth]{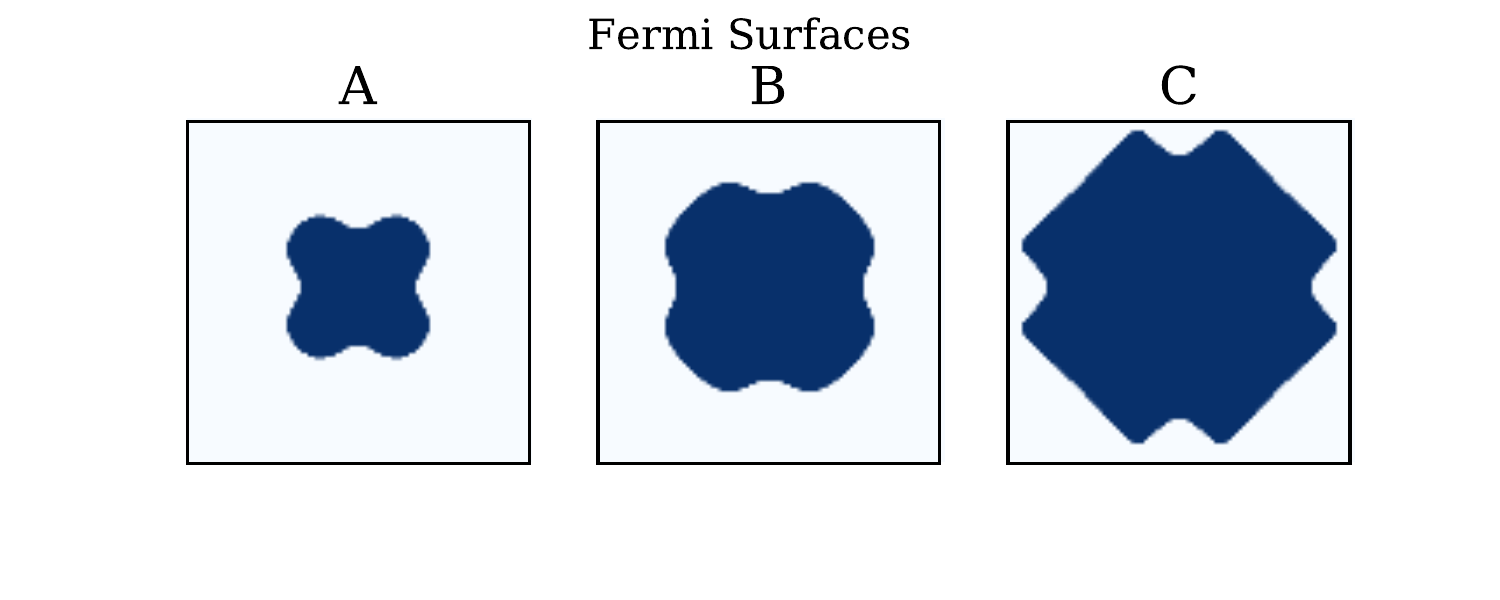}
  \includegraphics[width=0.45\textwidth]{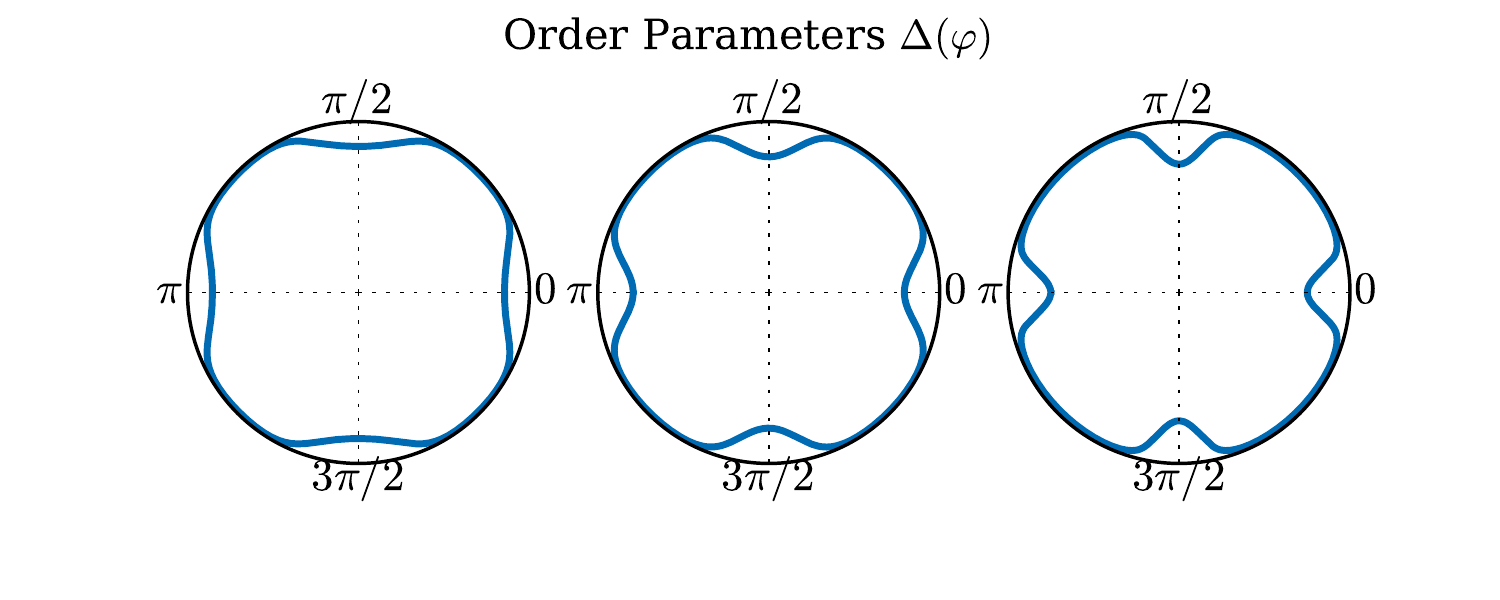}
  \caption{(Color online) Cooper pairing of fermions in shaken square lattice.
    TOP: Superfluid critical temperature $k_BT_c$ in units of band-width
    $W\equiv\max\epsilon_1-\min\epsilon_1$ for lattice depth $V_0=5E_R$,
    shaking parameters $s_0=0.05$, $\omega=1.01\Omega$, and interaction
    $|U|=0.2W$. The pairing order parameter has $s+d$-wave symmetry within the
    gray shaded region, and is predominantly $s$-wave otherwise. MIDDLE: The
    Fermi surfaces for points $A$, $B$ and $C$. BOTTOM: Angular dependence of
    order parameter $\Delta(\varphi)$ around the Fermi surface.}
  \label{fig:phase_diagram}
    %Onsite interaction constant is chosen as $U/W=0.3$ here.}
\end{figure}

Due to the momentum dependence of $V_\mathrm{eff}(\kb,\kb')$, the solution to 
Eq.~\eqref{eq:eigenvalue_equation} will yield an order parameter $\Delta(\kb)$
that depends on the angular location $\varphi$ of $\kb$ on the Fermi surface.
In other words, $\Delta(\kb)$ is in general anisotropic and includes 
higher harmonics,
\begin{equation}
\Delta(\varphi)=\Delta_s+\Delta_d\cos(2\varphi)+...
\end{equation}
Fig.~\ref{fig:phase_diagram} shows the phase diagram of $H_\mathrm{BCS}$ as
functions of filling and temperature.  In the gray shaded region, we find the
pairing is predominantly of $s+d$-wave symmetry, i.e., with a small $\Delta_d$
component and higher harmonics can be neglected.  This is illustrated for
three different fillings marked as $A$, $B$ and $C$ in the top row of
Fig.~\ref{fig:phase_diagram}.  The corresponding Fermi surfaces are shown in
the middle row, and the order parameters $\Delta(\varphi)$ are shown in the
bottom row.  The reduction in pairing amplitude along the $\Gamma \mathrm{X}$
line is consistent with the weaker effective interaction there found earlier
in section \ref{sec:interactions}. Thus the $s+d$-wave pairing is a direct
result of shaken 
induced anisotropy of the effective interactions. 
For other fillings outside the shaded region, the pairing symmetry is 
the usual $s$-wave.

%}}}
\section{Red-detuned Near-resonance Shaking}%{{{
\label{sec:resonant}
\begin{figure}[t]
  \centering
  \includegraphics[width=0.42\textwidth]{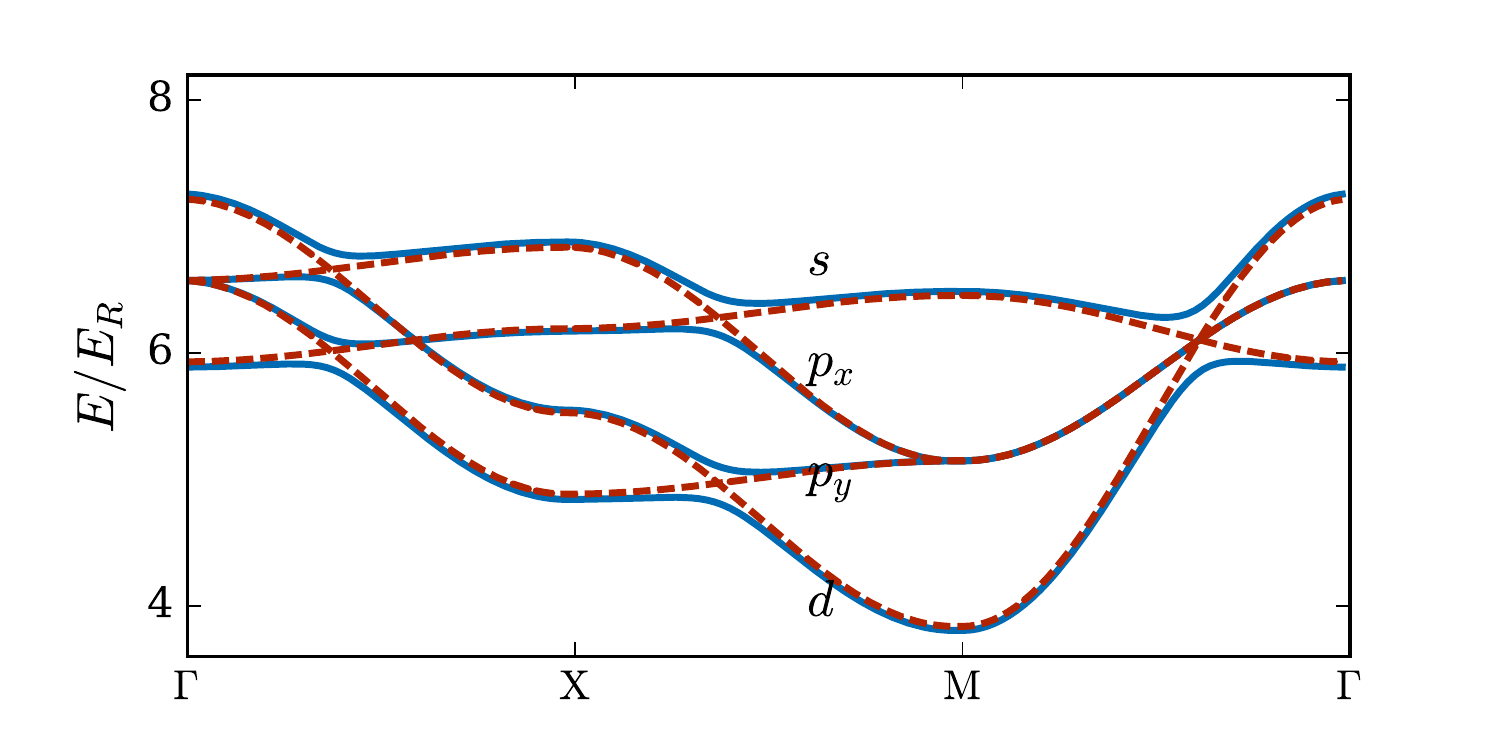}
  \includegraphics[width=0.42\textwidth]{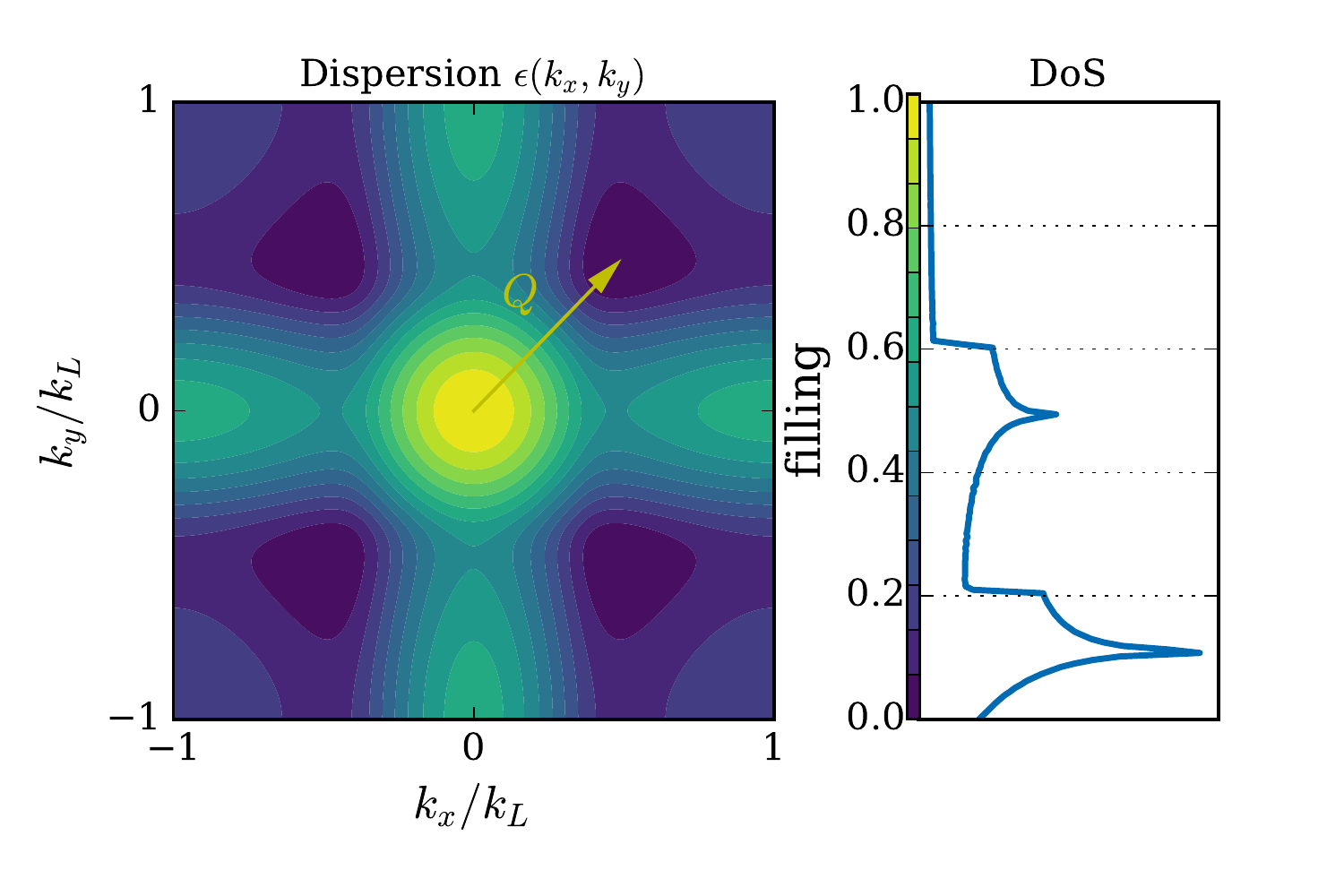}
  \includegraphics[width=0.42\textwidth]{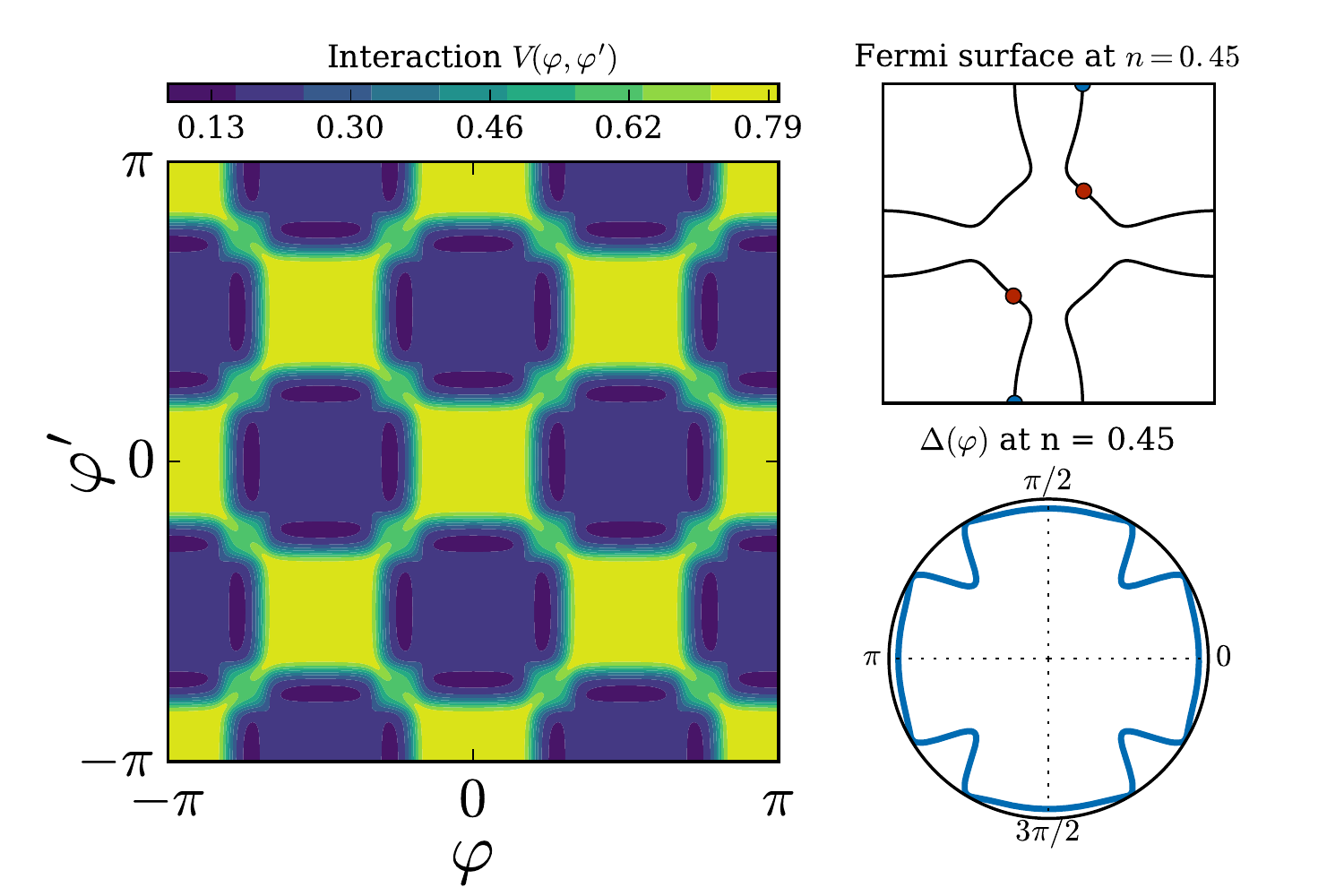}
  \caption{(Color online) Interacting fermions in shaken square lattice with
    red-detuned shaking frequency. TOP: Band mixing between the lowest four
    orbitals via
    shaking along the high symmetry points. Dashed (red) lines are bare
    energies whereas solid (blue) lines are shaking hybridized energies.
    MIDDLE: Dispersion of the hybridized $s$-band in the full Brillouin zone
    on the left and the corresponding density of states at the Fermi level for
    given filling on the right. BOTTOM: The angular dependence of interaction
    on the Fermi surface on the left for the Fermi surface shown on the right.
    Mean field order parameter around the Fermi surface with respect to center
    $\mathrm{M}$-point is shown below the Fermi surface plot. 
    Parameters used are $e_s=-0.56, t_s= -0.07,
    e_p= 2.76, t_p=0.42, d_0=2.63, d_1=-0.28$ corresponding to $V_0/E_R=5$.
    Shaking frequency is taken as $\omega = 0.85\Omega$ whereas the amplitude
    is $\beta=0.05$. }
  \label{fig:small_omega}
\end{figure}

So far we have focused on shaking frequencies near resonance, but blue detuned
from the band separation (not the band gap) of the $s$- and $p$-bands at the
Brillouin zone center, $\Omega = \max E_{p}-\min E_s$.  For example, we have
set $\omega/ \Omega=1.01$ and $1.05$. Now we move on to discuss red detuned
shaking frequencies $\omega<\Omega$, e.g., $\omega/ \Omega=0.85$, and compare
them to the blue detuned case.  For red detuned frequencies, the folded $s$-
and $p$-bands directly cross each other (see the dashed lines in the top panel
of Fig.~\ref{fig:small_omega}), and hybridize strongly near these crossing
points located away from the Brillouin zone center.
We emphasize that the theory developed in the previous sections are valid for
all near resonance shaking frequencies, $\omega\sim \Omega$.  The expression
for $H_\mathrm{eff}$, the band dispersion in
Eq.~\eqref{eq:effective_dispersion}, and the effective interactions in
Eq.~\eqref{eq:full_interaction} can be directly applied to red detuned cases
without any change. 

The calculation for $\omega/ \Omega=0.85$ proceeds the same way as before, and
the results are summarized in Fig.~\ref{fig:small_omega}.
A main difference from the blue detuned case  is the orbital character of the
hybridized $s$-band, shown at the top row of Fig. \ref{fig:dispersion} with
solid line.  It is predominantly of $d_{xy}$-orbital character at the
Brillouin zone center $\Gamma$, becomes more $p$-orbital like on the side of
the Brillouin zone around $\mathrm{X}$, and remains $s$-orbital like at the
Brillouin zone corner $\mathrm{M}$.  The shaking induced band hybridization is
strongest near the crossing points of the folded bare bands (dashed lines). As
shown in the middle panel of Fig.~\ref{fig:small_omega}, the dispersion of the
hybridized $s$-band also has four minima.  Compared to the blue detuned
case, the location of the energy minima, characterized by momentum vector
$\mathbf{Q}$, is further away from $\Gamma$ even though a smaller shaking
amplitude $\beta=0.05$ is used. The shapes of the Fermi surfaces  are quite
different as indicated by the contour lines of $\epsilon ({k_x,k_y})$. For
example, at filling $n=0.45$, the Fermi surface is centered at the Brillouin
zone corner $M$. 

In the red detuned regime, the anisotropy of effective interactions on the
Fermi surface is much more pronounced.  Take again $n=0.45$ as an example.
The effective interaction in the pairing channel $V(\varphi,\varphi')$ varies
by as much as seven folds. This is because the hybridized $s$-band contains
a significant contribution from the $d$-band, the bare interaction of which is
smaller than those of the $s$- and $p$-bands. As a result, $V$ is smaller for
$\mathbf{k}$ points closer to $\Gamma$, e.g.  the red dots in
Fig.~\ref{fig:small_omega}, and thus of more $d$-orbital character.
Due to the anisotropy of $V$, the pairing order parameter has $s+d$-wave
symmetry, as shown in the bottom panel of Fig.~\ref{fig:small_omega}. It
differs slightly from the one given in Fig.~\ref{fig:phase_diagram} in two
aspects. First, the $d$-wave component of the order parameter is increased.
Secondly, $\Delta(\varphi)$ reaches minimum at $\varphi=\pi/4$ where the
effective interaction is weak (recall again that in this case, the center of
Fermi surface is at $M$). 
%}}}
\section{Concluding remarks}%{{{
\label{sec:conclude}

Lattice shaking provides a valuable tool to engineer the band structures and
effective interactions for cold atoms in optical lattice beyond the reach of
the static optical lattices. The key physics at play is the mixing or
hybridization of different orbital bands induced by lattice shaking.
Presently, the new phases arising from interactions in these mixed bands
remain largely unexplored.  Our work constitutes a first step towards a
quantitative understanding of the interaction effects for fermionic atoms in
shaken square optical lattice. We derived a four-band effective Hamiltonian to
clarify the matrix elements for inter-orbital couplings and obtained the
analytical expressions for the eigenenergy bands, in good agreement with the
full numerical Floquet analysis. We further derived the effective interactions
on the hybridized $s$-band and explained the origin of its acquired momentum
dependence. Applying the theory to spin 1/2 fermions with attractive
interactions, we monitored the nontrivial evolution of the Fermi surface and
found the symmetry of the pairing order parameter can be $s+d$ wave. The
similarities and differences between the red and blue detuned shaking
frequency are discussed using insights gained from the analytical
understanding of $H_\mathrm{eff}$ and $V_\mathrm{eff}(\kb,\kb')$. 
These concrete examples support the presence of complex Fermi surfaces,
anisotropic interactions, and interesting many-body phases for fermions in
shaken lattices.

Our work can be generalized in several directions. For example, for repulsive
interactions, the general effective interaction vertex can be obtained and
subsequently applied to discuss competing many-body phases including spin
density waves, superfluidity, and Pomeranchuk instability etc. In particular,
interesting phenomena are expected when the Fermi surface are partially
nested, i.e., with large segments connected by some common nesting wave
vector. The technical procedures outlined here to derive the effective
Hamiltonian will also be useful to treat other regimes of shaking frequencies,
such as two-photon resonances known to give rise topologically nontrivial
bands.  Lastly, another interesting generalization is to analyze other lattice
geometries, e.g., interacting fermions on the shaken honeycomb and
checkerboard lattice. 
%}}}
\begin{acknowledgments}%{{{
We thank C. Chin for helpful discussions. This work is supported by AFOSR
Grant No.
FA9550-16-1-0006 (A.K., E.Z., and W.V.L.), ARO Grant No. W911NF-11-1-0230
(A.K. and W.V.L.), NSF Grant No. PHY-1205504 (A.K. and E.Z.), and NSF of China
Overseas Scholar Collaborative Program Grant No. 11429402 sponsored by Peking
University (W.V.L.).
\end{acknowledgments}
%}}}
\bibliography{refs}

\begin{thebibliography}{24}%
\makeatletter
\providecommand \@ifxundefined [1]{%
 \@ifx{#1\undefined}
}%
\providecommand \@ifnum [1]{%
 \ifnum #1\expandafter \@firstoftwo
 \else \expandafter \@secondoftwo
 \fi
}%
\providecommand \@ifx [1]{%
 \ifx #1\expandafter \@firstoftwo
 \else \expandafter \@secondoftwo
 \fi
}%
\providecommand \natexlab [1]{#1}%
\providecommand \enquote  [1]{``#1''}%
\providecommand \bibnamefont  [1]{#1}%
\providecommand \bibfnamefont [1]{#1}%
\providecommand \citenamefont [1]{#1}%
\providecommand \href@noop [0]{\@secondoftwo}%
\providecommand \href [0]{\begingroup \@sanitize@url \@href}%
\providecommand \@href[1]{\@@startlink{#1}\@@href}%
\providecommand \@@href[1]{\endgroup#1\@@endlink}%
\providecommand \@sanitize@url [0]{\catcode `\\12\catcode `\$12\catcode
  `\&12\catcode `\#12\catcode `\^12\catcode `\_12\catcode `\%12\relax}%
\providecommand \@@startlink[1]{}%
\providecommand \@@endlink[0]{}%
\providecommand \url  [0]{\begingroup\@sanitize@url \@url }%
\providecommand \@url [1]{\endgroup\@href {#1}{\urlprefix }}%
\providecommand \urlprefix  [0]{URL }%
\providecommand \Eprint [0]{\href }%
\providecommand \doibase [0]{http://dx.doi.org/}%
\providecommand \selectlanguage [0]{\@gobble}%
\providecommand \bibinfo  [0]{\@secondoftwo}%
\providecommand \bibfield  [0]{\@secondoftwo}%
\providecommand \translation [1]{[#1]}%
\providecommand \BibitemOpen [0]{}%
\providecommand \bibitemStop [0]{}%
\providecommand \bibitemNoStop [0]{.\EOS\space}%
\providecommand \EOS [0]{\spacefactor3000\relax}%
\providecommand \BibitemShut  [1]{\csname bibitem#1\endcsname}%
\let\auto@bib@innerbib\@empty
%</preamble>
\bibitem [{\citenamefont {Madison}\ \emph {et~al.}(1998)\citenamefont
  {Madison}, \citenamefont {Fischer}, \citenamefont {Diener}, \citenamefont
  {Niu},\ and\ \citenamefont {Raizen}}]{PhysRevLett.81.5093}%
  \BibitemOpen
  \bibfield  {author} {\bibinfo {author} {\bibfnamefont {K.~W.}\ \bibnamefont
  {Madison}}, \bibinfo {author} {\bibfnamefont {M.~C.}\ \bibnamefont
  {Fischer}}, \bibinfo {author} {\bibfnamefont {R.~B.}\ \bibnamefont {Diener}},
  \bibinfo {author} {\bibfnamefont {Qian}\ \bibnamefont {Niu}}, \ and\ \bibinfo
  {author} {\bibfnamefont {M.~G.}\ \bibnamefont {Raizen}},\ }\bibfield  {title}
  {\enquote {\bibinfo {title} {Dynamical bloch band suppression in an optical
  lattice},}\ }\href {\doibase 10.1103/PhysRevLett.81.5093} {\bibfield
  {journal} {\bibinfo  {journal} {Phys. Rev. Lett.}\ }\textbf {\bibinfo
  {volume} {81}},\ \bibinfo {pages} {5093--5096} (\bibinfo {year}
  {1998})}\BibitemShut {NoStop}%
\bibitem [{\citenamefont {Lignier}\ \emph {et~al.}(2007)\citenamefont
  {Lignier}, \citenamefont {Sias}, \citenamefont {Ciampini}, \citenamefont
  {Singh}, \citenamefont {Zenesini}, \citenamefont {Morsch},\ and\
  \citenamefont {Arimondo}}]{PhysRevLett.99.220403}%
  \BibitemOpen
  \bibfield  {author} {\bibinfo {author} {\bibfnamefont {H.}~\bibnamefont
  {Lignier}}, \bibinfo {author} {\bibfnamefont {C.}~\bibnamefont {Sias}},
  \bibinfo {author} {\bibfnamefont {D.}~\bibnamefont {Ciampini}}, \bibinfo
  {author} {\bibfnamefont {Y.}~\bibnamefont {Singh}}, \bibinfo {author}
  {\bibfnamefont {A.}~\bibnamefont {Zenesini}}, \bibinfo {author}
  {\bibfnamefont {O.}~\bibnamefont {Morsch}}, \ and\ \bibinfo {author}
  {\bibfnamefont {E.}~\bibnamefont {Arimondo}},\ }\bibfield  {title} {\enquote
  {\bibinfo {title} {Dynamical control of matter-wave tunneling in periodic
  potentials},}\ }\href {\doibase 10.1103/PhysRevLett.99.220403} {\bibfield
  {journal} {\bibinfo  {journal} {Phys. Rev. Lett.}\ }\textbf {\bibinfo
  {volume} {99}},\ \bibinfo {pages} {220403} (\bibinfo {year}
  {2007})}\BibitemShut {NoStop}%
\bibitem [{\citenamefont {Zenesini}\ \emph {et~al.}(2009)\citenamefont
  {Zenesini}, \citenamefont {Lignier}, \citenamefont {Ciampini}, \citenamefont
  {Morsch},\ and\ \citenamefont {Arimondo}}]{Zenesini2009}%
  \BibitemOpen
  \bibfield  {author} {\bibinfo {author} {\bibfnamefont {Alessandro}\
  \bibnamefont {Zenesini}}, \bibinfo {author} {\bibfnamefont {Hans}\
  \bibnamefont {Lignier}}, \bibinfo {author} {\bibfnamefont {Donatella}\
  \bibnamefont {Ciampini}}, \bibinfo {author} {\bibfnamefont {Oliver}\
  \bibnamefont {Morsch}}, \ and\ \bibinfo {author} {\bibfnamefont {Ennio}\
  \bibnamefont {Arimondo}},\ }\bibfield  {title} {\enquote {\bibinfo {title}
  {{Coherent Control of Dressed Matter Waves}},}\ }\href {\doibase
  10.1103/PhysRevLett.102.100403} {\bibfield  {journal} {\bibinfo  {journal}
  {Phys. Rev. Lett.}\ }\textbf {\bibinfo {volume} {102}},\ \bibinfo {pages}
  {100403} (\bibinfo {year} {2009})}\BibitemShut {NoStop}%
\bibitem [{\citenamefont {Holthaus}(2016)}]{Holthaus2015a}%
  \BibitemOpen
  \bibfield  {author} {\bibinfo {author} {\bibfnamefont {Martin}\ \bibnamefont
  {Holthaus}},\ }\bibfield  {title} {\enquote {\bibinfo {title} {{Floquet
  engineering with quasienergy bands of periodically driven optical
  lattices}},}\ }\href {\doibase 10.1088/0953-4075/49/1/013001} {\bibfield
  {journal} {\bibinfo  {journal} {J. Phys. B At. Mol. Opt. Phys.}\ }\textbf
  {\bibinfo {volume} {49}},\ \bibinfo {pages} {13001} (\bibinfo {year}
  {2016})}\BibitemShut {NoStop}%
\bibitem [{\citenamefont {Goldman}\ \emph {et~al.}(2014)\citenamefont
  {Goldman}, \citenamefont {Juzeliūnas}, \citenamefont {{\"{O}}hberg},\ and\
  \citenamefont {Spielman}}]{Goldman2014a}%
  \BibitemOpen
  \bibfield  {author} {\bibinfo {author} {\bibfnamefont {N}~\bibnamefont
  {Goldman}}, \bibinfo {author} {\bibfnamefont {G}~\bibnamefont {Juzeliūnas}},
  \bibinfo {author} {\bibfnamefont {P}~\bibnamefont {{\"{O}}hberg}}, \ and\
  \bibinfo {author} {\bibfnamefont {I~B}\ \bibnamefont {Spielman}},\ }\bibfield
   {title} {\enquote {\bibinfo {title} {{Light-induced gauge fields for
  ultracold atoms}},}\ }\href {\doibase 10.1088/0034-4885/77/12/126401}
  {\bibfield  {journal} {\bibinfo  {journal} {Reports Prog. Phys.}\ }\textbf
  {\bibinfo {volume} {77}},\ \bibinfo {pages} {126401} (\bibinfo {year}
  {2014})}\BibitemShut {NoStop}%
\bibitem [{\citenamefont {Eckardt}(2016)}]{eckardt2016atomic}%
  \BibitemOpen
  \bibfield  {author} {\bibinfo {author} {\bibfnamefont {Andr{\'e}}\
  \bibnamefont {Eckardt}},\ }\bibfield  {title} {\enquote {\bibinfo {title}
  {Atomic quantum gases in periodically driven optical lattices},}\ }\href
  {http://arxiv.org/abs/1606.08041} {\bibfield  {journal} {\bibinfo  {journal}
  {arXiv:1606.08041}\ } (\bibinfo {year} {2016})}\BibitemShut {NoStop}%
\bibitem [{\citenamefont {Struck}\ \emph {et~al.}(2011)\citenamefont {Struck},
  \citenamefont {{\"O}lschl{\"a}ger}, \citenamefont {Le~Targat}, \citenamefont
  {Soltan-Panahi}, \citenamefont {Eckardt}, \citenamefont {Lewenstein},
  \citenamefont {Windpassinger},\ and\ \citenamefont {Sengstock}}]{Struck2011}%
  \BibitemOpen
  \bibfield  {author} {\bibinfo {author} {\bibfnamefont {Julian}\ \bibnamefont
  {Struck}}, \bibinfo {author} {\bibfnamefont {Christoph}\ \bibnamefont
  {{\"O}lschl{\"a}ger}}, \bibinfo {author} {\bibfnamefont {R}~\bibnamefont
  {Le~Targat}}, \bibinfo {author} {\bibfnamefont {Parvis}\ \bibnamefont
  {Soltan-Panahi}}, \bibinfo {author} {\bibfnamefont {Andr{\'e}}\ \bibnamefont
  {Eckardt}}, \bibinfo {author} {\bibfnamefont {Maciej}\ \bibnamefont
  {Lewenstein}}, \bibinfo {author} {\bibfnamefont {Patrick}\ \bibnamefont
  {Windpassinger}}, \ and\ \bibinfo {author} {\bibfnamefont {Klaus}\
  \bibnamefont {Sengstock}},\ }\bibfield  {title} {\enquote {\bibinfo {title}
  {{Quantum simulation of frustrated classical magnetism in triangular optical
  lattices.}}}\ }\href {\doibase 10.1126/science.1207239} {\bibfield  {journal}
  {\bibinfo  {journal} {Science}\ }\textbf {\bibinfo {volume} {333}},\ \bibinfo
  {pages} {996--9} (\bibinfo {year} {2011})}\BibitemShut {NoStop}%
\bibitem [{\citenamefont {Struck}\ \emph {et~al.}(2012)\citenamefont {Struck},
  \citenamefont {{\"{O}}lschl{\"{a}}ger}, \citenamefont {Weinberg},
  \citenamefont {Hauke}, \citenamefont {Simonet}, \citenamefont {Eckardt},
  \citenamefont {Lewenstein}, \citenamefont {Sengstock},\ and\ \citenamefont
  {Windpassinger}}]{Struck2012}%
  \BibitemOpen
  \bibfield  {author} {\bibinfo {author} {\bibfnamefont {J}~\bibnamefont
  {Struck}}, \bibinfo {author} {\bibfnamefont {C}~\bibnamefont
  {{\"{O}}lschl{\"{a}}ger}}, \bibinfo {author} {\bibfnamefont {M}~\bibnamefont
  {Weinberg}}, \bibinfo {author} {\bibfnamefont {P}~\bibnamefont {Hauke}},
  \bibinfo {author} {\bibfnamefont {J}~\bibnamefont {Simonet}}, \bibinfo
  {author} {\bibfnamefont {A}~\bibnamefont {Eckardt}}, \bibinfo {author}
  {\bibfnamefont {M}~\bibnamefont {Lewenstein}}, \bibinfo {author}
  {\bibfnamefont {K}~\bibnamefont {Sengstock}}, \ and\ \bibinfo {author}
  {\bibfnamefont {P}~\bibnamefont {Windpassinger}},\ }\bibfield  {title}
  {\enquote {\bibinfo {title} {{Tunable gauge potential for neutral and
  spinless particles in driven optical lattices.}}}\ }\href {\doibase
  10.1103/PhysRevLett.108.225304} {\bibfield  {journal} {\bibinfo  {journal}
  {Phys. Rev. Lett.}\ }\textbf {\bibinfo {volume} {108}},\ \bibinfo {pages}
  {225304} (\bibinfo {year} {2012})}\BibitemShut {NoStop}%
\bibitem [{\citenamefont {Hauke}\ \emph {et~al.}(2012)\citenamefont {Hauke},
  \citenamefont {Tieleman}, \citenamefont {Celi}, \citenamefont
  {{\"{O}}lschl{\"{a}}ger}, \citenamefont {Simonet}, \citenamefont {Struck},
  \citenamefont {Weinberg}, \citenamefont {Windpassinger}, \citenamefont
  {Sengstock}, \citenamefont {Lewenstein},\ and\ \citenamefont
  {Eckardt}}]{Hauke2012b}%
  \BibitemOpen
  \bibfield  {author} {\bibinfo {author} {\bibfnamefont {Philipp}\ \bibnamefont
  {Hauke}}, \bibinfo {author} {\bibfnamefont {Olivier}\ \bibnamefont
  {Tieleman}}, \bibinfo {author} {\bibfnamefont {Alessio}\ \bibnamefont
  {Celi}}, \bibinfo {author} {\bibfnamefont {Christoph}\ \bibnamefont
  {{\"{O}}lschl{\"{a}}ger}}, \bibinfo {author} {\bibfnamefont {Juliette}\
  \bibnamefont {Simonet}}, \bibinfo {author} {\bibfnamefont {Julian}\
  \bibnamefont {Struck}}, \bibinfo {author} {\bibfnamefont {Malte}\
  \bibnamefont {Weinberg}}, \bibinfo {author} {\bibfnamefont {Patrick}\
  \bibnamefont {Windpassinger}}, \bibinfo {author} {\bibfnamefont {Klaus}\
  \bibnamefont {Sengstock}}, \bibinfo {author} {\bibfnamefont {Maciej}\
  \bibnamefont {Lewenstein}}, \ and\ \bibinfo {author} {\bibfnamefont
  {Andr{\'{e}}}\ \bibnamefont {Eckardt}},\ }\bibfield  {title} {\enquote
  {\bibinfo {title} {{Non-Abelian Gauge Fields and Topological Insulators in
  Shaken Optical Lattices}},}\ }\href {\doibase 10.1103/PhysRevLett.109.145301}
  {\bibfield  {journal} {\bibinfo  {journal} {Phys. Rev. Lett.}\ }\textbf
  {\bibinfo {volume} {109}},\ \bibinfo {pages} {145301} (\bibinfo {year}
  {2012})}\BibitemShut {NoStop}%
\bibitem [{\citenamefont {Struck}\ \emph {et~al.}(2013)\citenamefont {Struck},
  \citenamefont {Weinberg}, \citenamefont {{\"{O}}lschl{\"{a}}ger},
  \citenamefont {Windpassinger}, \citenamefont {Simonet}, \citenamefont
  {Sengstock}, \citenamefont {H{\"{o}}ppner}, \citenamefont {Hauke},
  \citenamefont {Eckardt}, \citenamefont {Lewenstein},\ and\ \citenamefont
  {Mathey}}]{Struck2013}%
  \BibitemOpen
  \bibfield  {author} {\bibinfo {author} {\bibfnamefont {J.}~\bibnamefont
  {Struck}}, \bibinfo {author} {\bibfnamefont {M.}~\bibnamefont {Weinberg}},
  \bibinfo {author} {\bibfnamefont {C.}~\bibnamefont {{\"{O}}lschl{\"{a}}ger}},
  \bibinfo {author} {\bibfnamefont {P.}~\bibnamefont {Windpassinger}}, \bibinfo
  {author} {\bibfnamefont {J.}~\bibnamefont {Simonet}}, \bibinfo {author}
  {\bibfnamefont {K.}~\bibnamefont {Sengstock}}, \bibinfo {author}
  {\bibfnamefont {R.}~\bibnamefont {H{\"{o}}ppner}}, \bibinfo {author}
  {\bibfnamefont {P.}~\bibnamefont {Hauke}}, \bibinfo {author} {\bibfnamefont
  {A.}~\bibnamefont {Eckardt}}, \bibinfo {author} {\bibfnamefont
  {M.}~\bibnamefont {Lewenstein}}, \ and\ \bibinfo {author} {\bibfnamefont
  {L.}~\bibnamefont {Mathey}},\ }\bibfield  {title} {\enquote {\bibinfo {title}
  {{Engineering Ising-XY spin-models in a triangular lattice using tunable
  artificial gauge fields}},}\ }\href {\doibase 10.1038/nphys2750} {\bibfield
  {journal} {\bibinfo  {journal} {Nat. Phys.}\ }\textbf {\bibinfo {volume}
  {9}},\ \bibinfo {pages} {738--743} (\bibinfo {year} {2013})}\BibitemShut
  {NoStop}%
\bibitem [{\citenamefont {Jotzu}\ \emph {et~al.}(2014)\citenamefont {Jotzu},
  \citenamefont {Messer}, \citenamefont {Desbuquois}, \citenamefont {Lebrat},
  \citenamefont {Uehlinger}, \citenamefont {Greif},\ and\ \citenamefont
  {Esslinger}}]{Jotzu2014}%
  \BibitemOpen
  \bibfield  {author} {\bibinfo {author} {\bibfnamefont {Gregor}\ \bibnamefont
  {Jotzu}}, \bibinfo {author} {\bibfnamefont {Michael}\ \bibnamefont {Messer}},
  \bibinfo {author} {\bibfnamefont {R{\'{e}}mi}\ \bibnamefont {Desbuquois}},
  \bibinfo {author} {\bibfnamefont {Martin}\ \bibnamefont {Lebrat}}, \bibinfo
  {author} {\bibfnamefont {Thomas}\ \bibnamefont {Uehlinger}}, \bibinfo
  {author} {\bibfnamefont {Daniel}\ \bibnamefont {Greif}}, \ and\ \bibinfo
  {author} {\bibfnamefont {Tilman}\ \bibnamefont {Esslinger}},\ }\bibfield
  {title} {\enquote {\bibinfo {title} {{Experimental realization of the
  topological Haldane model with ultracold fermions}},}\ }\href {\doibase
  10.1038/nature13915} {\bibfield  {journal} {\bibinfo  {journal} {Nature}\
  }\textbf {\bibinfo {volume} {515}},\ \bibinfo {pages} {237--240} (\bibinfo
  {year} {2014})}\BibitemShut {NoStop}%
\bibitem [{\citenamefont {Parker}\ \emph {et~al.}(2013)\citenamefont {Parker},
  \citenamefont {Ha},\ and\ \citenamefont {Chin}}]{Parker2013}%
  \BibitemOpen
  \bibfield  {author} {\bibinfo {author} {\bibfnamefont {Colin~V.}\
  \bibnamefont {Parker}}, \bibinfo {author} {\bibfnamefont {Li-Chung}\
  \bibnamefont {Ha}}, \ and\ \bibinfo {author} {\bibfnamefont {Cheng}\
  \bibnamefont {Chin}},\ }\bibfield  {title} {\enquote {\bibinfo {title}
  {{Direct observation of effective ferromagnetic domains of cold atoms in a
  shaken optical lattice}},}\ }\href {\doibase 10.1038/nphys2789} {\bibfield
  {journal} {\bibinfo  {journal} {Nat. Phys.}\ }\textbf {\bibinfo {volume}
  {9}},\ \bibinfo {pages} {769--774} (\bibinfo {year} {2013})}\BibitemShut
  {NoStop}%
\bibitem [{\citenamefont {Clark}\ \emph {et~al.}(2016)\citenamefont {Clark},
  \citenamefont {Feng},\ and\ \citenamefont {Chin}}]{Clark2016}%
  \BibitemOpen
  \bibfield  {author} {\bibinfo {author} {\bibfnamefont {Logan~W.}\
  \bibnamefont {Clark}}, \bibinfo {author} {\bibfnamefont {Lei}\ \bibnamefont
  {Feng}}, \ and\ \bibinfo {author} {\bibfnamefont {Cheng}\ \bibnamefont
  {Chin}},\ }\bibfield  {title} {\enquote {\bibinfo {title} {{Universal
  space-time scaling symmetry in the dynamics of bosons across a quantum phase
  transition}},}\ }\href {\doibase 10.1126/science.aaf9657} {\bibfield
  {journal} {\bibinfo  {journal} {Science}\ }\textbf {\bibinfo {volume}
  {354}},\ \bibinfo {pages} {606--610} (\bibinfo {year} {2016})}\BibitemShut
  {NoStop}%
\bibitem [{\citenamefont {Eckardt}\ \emph {et~al.}(2005)\citenamefont
  {Eckardt}, \citenamefont {Weiss},\ and\ \citenamefont
  {Holthaus}}]{PhysRevLett.95.260404}%
  \BibitemOpen
  \bibfield  {author} {\bibinfo {author} {\bibfnamefont {Andr\'e}\ \bibnamefont
  {Eckardt}}, \bibinfo {author} {\bibfnamefont {Christoph}\ \bibnamefont
  {Weiss}}, \ and\ \bibinfo {author} {\bibfnamefont {Martin}\ \bibnamefont
  {Holthaus}},\ }\bibfield  {title} {\enquote {\bibinfo {title}
  {Superfluid-insulator transition in a periodically driven optical lattice},}\
  }\href {\doibase 10.1103/PhysRevLett.95.260404} {\bibfield  {journal}
  {\bibinfo  {journal} {Phys. Rev. Lett.}\ }\textbf {\bibinfo {volume} {95}},\
  \bibinfo {pages} {260404} (\bibinfo {year} {2005})}\BibitemShut {NoStop}%
\bibitem [{\citenamefont {S\o{}rensen}\ \emph {et~al.}(2005)\citenamefont
  {S\o{}rensen}, \citenamefont {Demler},\ and\ \citenamefont
  {Lukin}}]{PhysRevLett.94.086803}%
  \BibitemOpen
  \bibfield  {author} {\bibinfo {author} {\bibfnamefont {Anders~S.}\
  \bibnamefont {S\o{}rensen}}, \bibinfo {author} {\bibfnamefont {Eugene}\
  \bibnamefont {Demler}}, \ and\ \bibinfo {author} {\bibfnamefont {Mikhail~D.}\
  \bibnamefont {Lukin}},\ }\bibfield  {title} {\enquote {\bibinfo {title}
  {Fractional quantum hall states of atoms in optical lattices},}\ }\href
  {\doibase 10.1103/PhysRevLett.94.086803} {\bibfield  {journal} {\bibinfo
  {journal} {Phys. Rev. Lett.}\ }\textbf {\bibinfo {volume} {94}},\ \bibinfo
  {pages} {086803} (\bibinfo {year} {2005})}\BibitemShut {NoStop}%
\bibitem [{\citenamefont {Goldman}\ and\ \citenamefont
  {Dalibard}(2014)}]{PhysRevX.4.031027}%
  \BibitemOpen
  \bibfield  {author} {\bibinfo {author} {\bibfnamefont {N.}~\bibnamefont
  {Goldman}}\ and\ \bibinfo {author} {\bibfnamefont {J.}~\bibnamefont
  {Dalibard}},\ }\bibfield  {title} {\enquote {\bibinfo {title} {Periodically
  driven quantum systems: Effective hamiltonians and engineered gauge
  fields},}\ }\href {\doibase 10.1103/PhysRevX.4.031027} {\bibfield  {journal}
  {\bibinfo  {journal} {Phys. Rev. X}\ }\textbf {\bibinfo {volume} {4}},\
  \bibinfo {pages} {031027} (\bibinfo {year} {2014})}\BibitemShut {NoStop}%
\bibitem [{\citenamefont {Zhang}\ \emph {et~al.}(2015)\citenamefont {Zhang},
  \citenamefont {Lang},\ and\ \citenamefont {Zhou}}]{Zhang2015a}%
  \BibitemOpen
  \bibfield  {author} {\bibinfo {author} {\bibfnamefont {Shao-Liang}\
  \bibnamefont {Zhang}}, \bibinfo {author} {\bibfnamefont {Li-Jun}\
  \bibnamefont {Lang}}, \ and\ \bibinfo {author} {\bibfnamefont
  {Qi}~\bibnamefont {Zhou}},\ }\bibfield  {title} {\enquote {\bibinfo {title}
  {{Chiral d-Wave Superfluid in Periodically Driven Lattices}},}\ }\href
  {\doibase 10.1103/PhysRevLett.115.225301} {\bibfield  {journal} {\bibinfo
  {journal} {Phys. Rev. Lett.}\ }\textbf {\bibinfo {volume} {115}},\ \bibinfo
  {pages} {225301} (\bibinfo {year} {2015})}\BibitemShut {NoStop}%
\bibitem [{\citenamefont {Sedrakyan}\ \emph {et~al.}(2015)\citenamefont
  {Sedrakyan}, \citenamefont {Galitski},\ and\ \citenamefont
  {Kamenev}}]{Sedrakyan2015}%
  \BibitemOpen
  \bibfield  {author} {\bibinfo {author} {\bibfnamefont {Tigran~A.}\
  \bibnamefont {Sedrakyan}}, \bibinfo {author} {\bibfnamefont {Victor~M.}\
  \bibnamefont {Galitski}}, \ and\ \bibinfo {author} {\bibfnamefont {Alex}\
  \bibnamefont {Kamenev}},\ }\bibfield  {title} {\enquote {\bibinfo {title}
  {{Statistical Transmutation in Floquet Driven Optical Lattices}},}\ }\href
  {\doibase 10.1103/PhysRevLett.115.195301} {\bibfield  {journal} {\bibinfo
  {journal} {Phys. Rev. Lett.}\ }\textbf {\bibinfo {volume} {115}},\ \bibinfo
  {pages} {195301} (\bibinfo {year} {2015})}\BibitemShut {NoStop}%
\bibitem [{\citenamefont {Zheng}\ and\ \citenamefont {Zhai}(2014)}]{Zheng2014}%
  \BibitemOpen
  \bibfield  {author} {\bibinfo {author} {\bibfnamefont {Wei}\ \bibnamefont
  {Zheng}}\ and\ \bibinfo {author} {\bibfnamefont {Hui}\ \bibnamefont {Zhai}},\
  }\bibfield  {title} {\enquote {\bibinfo {title} {{Floquet topological states
  in shaking optical lattices}},}\ }\href {\doibase 10.1103/PhysRevA.89.061603}
  {\bibfield  {journal} {\bibinfo  {journal} {Phys. Rev. A}\ }\textbf {\bibinfo
  {volume} {89}},\ \bibinfo {pages} {061603} (\bibinfo {year}
  {2014})}\BibitemShut {NoStop}%
\bibitem [{\citenamefont {Zhang}\ and\ \citenamefont {Zhou}(2014)}]{Zhang2014}%
  \BibitemOpen
  \bibfield  {author} {\bibinfo {author} {\bibfnamefont {Shao-Liang}\
  \bibnamefont {Zhang}}\ and\ \bibinfo {author} {\bibfnamefont
  {Qi}~\bibnamefont {Zhou}},\ }\bibfield  {title} {\enquote {\bibinfo {title}
  {{Shaping topological properties of the band structures in a shaken optical
  lattice}},}\ }\href {\doibase 10.1103/PhysRevA.90.051601} {\bibfield
  {journal} {\bibinfo  {journal} {Phys. Rev. A}\ }\textbf {\bibinfo {volume}
  {90}},\ \bibinfo {pages} {051601} (\bibinfo {year} {2014})}\BibitemShut
  {NoStop}%
\bibitem [{\citenamefont {Zheng}\ \emph {et~al.}(2015)\citenamefont {Zheng},
  \citenamefont {Qu}, \citenamefont {Zou},\ and\ \citenamefont
  {Zhang}}]{Zheng2015}%
  \BibitemOpen
  \bibfield  {author} {\bibinfo {author} {\bibfnamefont {Zhen}\ \bibnamefont
  {Zheng}}, \bibinfo {author} {\bibfnamefont {Chunlei}\ \bibnamefont {Qu}},
  \bibinfo {author} {\bibfnamefont {Xubo}\ \bibnamefont {Zou}}, \ and\ \bibinfo
  {author} {\bibfnamefont {Chuanwei}\ \bibnamefont {Zhang}},\ }\bibfield
  {title} {\enquote {\bibinfo {title} {{Floquet
  Fulde-Ferrell-Larkin-Ovchinnikov superfluids and Majorana fermions in a
  shaken fermionic optical lattice}},}\ }\href {\doibase
  10.1103/PhysRevA.91.063626} {\bibfield  {journal} {\bibinfo  {journal} {Phys.
  Rev. A}\ }\textbf {\bibinfo {volume} {91}},\ \bibinfo {pages} {063626}
  (\bibinfo {year} {2015})},\ \Eprint {http://arxiv.org/abs/1408.5824}
  {arXiv:1408.5824} \BibitemShut {NoStop}%
\bibitem [{\citenamefont {Miao}\ \emph {et~al.}(2015)\citenamefont {Miao},
  \citenamefont {Liu},\ and\ \citenamefont {Zheng}}]{Miao2015}%
  \BibitemOpen
  \bibfield  {author} {\bibinfo {author} {\bibfnamefont {Jiao}\ \bibnamefont
  {Miao}}, \bibinfo {author} {\bibfnamefont {Boyang}\ \bibnamefont {Liu}}, \
  and\ \bibinfo {author} {\bibfnamefont {Wei}\ \bibnamefont {Zheng}},\
  }\bibfield  {title} {\enquote {\bibinfo {title} {{Quantum phase transition of
  bosons in a shaken optical lattice}},}\ }\href {\doibase
  10.1103/PhysRevA.91.033404} {\bibfield  {journal} {\bibinfo  {journal}
  {Physical Review A}\ }\textbf {\bibinfo {volume} {91}},\ \bibinfo {pages}
  {033404} (\bibinfo {year} {2015})}\BibitemShut {NoStop}%
\bibitem [{\citenamefont {Po}\ and\ \citenamefont {Zhou}(2015)}]{Po2015}%
  \BibitemOpen
  \bibfield  {author} {\bibinfo {author} {\bibfnamefont {Hoi~Chun}\
  \bibnamefont {Po}}\ and\ \bibinfo {author} {\bibfnamefont {Qi}~\bibnamefont
  {Zhou}},\ }\bibfield  {title} {\enquote {\bibinfo {title} {{A two-dimensional
  algebraic quantum liquid produced by an atomic simulator of the quantum
  Lifshitz model}},}\ }\href {\doibase 10.1038/ncomms9012} {\bibfield
  {journal} {\bibinfo  {journal} {Nature Communications}\ }\textbf {\bibinfo
  {volume} {6}},\ \bibinfo {pages} {8012} (\bibinfo {year} {2015})}\BibitemShut
  {NoStop}%
\bibitem [{\citenamefont {Li}\ and\ \citenamefont {Liu}(2016)}]{Li2015}%
  \BibitemOpen
  \bibfield  {author} {\bibinfo {author} {\bibfnamefont {Xiaopeng}\
  \bibnamefont {Li}}\ and\ \bibinfo {author} {\bibfnamefont {W.~Vincent}\
  \bibnamefont {Liu}},\ }\bibfield  {title} {\enquote {\bibinfo {title}
  {{Physics of higher orbital bands in optical lattices: a review}},}\ }\href
  {\doibase 10.1088/0034-4885/79/11/116401} {\bibfield  {journal} {\bibinfo
  {journal} {Reports Prog. Phys.}\ }\textbf {\bibinfo {volume} {79}},\ \bibinfo
  {pages} {116401} (\bibinfo {year} {2016})}\BibitemShut {NoStop}%
\end{thebibliography}%

\end{document}